\documentclass[12 pt,twoside,aps,prd,amsmath,amssymb,
tightenlines,showpacs,showkeys]{revtex4-1}

\usepackage{amssymb}
\usepackage{amsmath}
\usepackage[dvips]{graphicx}
\usepackage{graphicx,epsfig}
\usepackage{bm}
\def\myfigure #1#2#3#4
{\begin{figure}[ht]\begin{center}
\includegraphics[width=#2 \textwidth]{#1.eps}
\parbox[t]{#4\textwidth}{\caption{#3}\label{#1}}
\end{center}\end{figure}}

\def \myfigures #1#2#3#4#5#6#7#8
{\begin{figure}[ht]
    \begin{center}
        \includegraphics[width=#2 \textwidth]{#1.eps}
        \hfill
        \includegraphics[width=#6 \textwidth]{#5.eps}
        \parbox[t]{#4\textwidth}{\caption {#3}\label{#1}}
        \hfill
        \parbox[t]{#8\textwidth}{\caption {#7}\label{#5}}
    \end{center}
\end{figure} }

\begin{document}
\baselineskip -24pt
\title{Bianchi type V bulk viscous cosmological models with particle creation in General Relativity}
\author{Mohd.Zeyauddin},
\affiliation{Bogoliubov Laboratory of Theoretical Physics\\
Joint Institute for Nuclear Research\\
141980 Dubna, Moscow region, Russia}
\email{zeya@theor.jinr.ru}\\\\
\author{Bijan Saha}
\affiliation{Laboratory of Information Technologies\\
Joint Institute for Nuclear Research\\
141980 Dubna, Moscow region, Russia} \email{bijan@jinr.ru}





\begin{abstract}
In this paper, a spatially homogeneous and anisotropic Bianchi type
V model filled with an imperfect fluid with bulk viscosity and
particle creation, is investigated within the framework of General
Relativity. Particle creation and bulk viscosity have been
considered as separate irreversible processes. The energy- momentum
tensor is modified to accommodate the viscous pressure and creation
pressure which is associated with creation of matter out of
gravitational field. Exact solutions of the field equations are
obtained by applying a special law of variation of Hubble parameter.
Using this assumption, we obtain two types of cosmological models.
We find a singularity in the first model whereas second model is
non-singular. Further we separately study the bulk viscosity and
particle creation in each model considering four different cases.
The bulk viscosity coefficient $\zeta$ is obtained for Truncated,
Full Causal and Eckart theories in all cases. All physical
parameters are calculated and thoroughly discussed in both models.
\end{abstract}


\maketitle

\bigskip

\pagebreak
\pagestyle{headings}
\section{Introduction}
Recently, particle creation processes are supposed to play an
important role in the early evolution of the universe.
Phenomenologically, particle creation has been described in terms of
effective bulk viscosity coefficients \cite{Zel, Hu, Lima, Zimdahl}.
Prigogine et al. \cite{Prigogine, Prigoginei} have investigated the
first theoretical approach of particle creation which was a
macroscopic phenomena allowing for both particle creation and
entropy production in the early universe. They have suggested that
matter creation take place out of gravitational energy in
irreversible process of the non-equilibrium thermodynamics. That
leads very naturally to a reinterpretation of the matter
stress-energy tensor in the Einstein's general relativity. This type
of creation basically corresponds to irreversible energy flow from
the gravitational field to the created matter constituent. The rate
of particle production can also be described by the quantum-field
theory in curved space-time \cite{Birrell}. The idea of elementary
particle creation in the expanding universe was given by Parker
\cite{Parker}. Hoyle and Narlikar \cite{Hoyle} had presented the
idea of continuous creation of matter which was subsequently
modified in the form of quasi-steady-state cosmology \cite{Hoylei}.
Eckart \cite{Eckart} developed the first relativistic theory of
non-equilibrium thermodynamics to study the effect of bulk
viscosity. Later on, it was pointed out that the Eckart's theory has
serious drawbacks concerning causality and stability. Gr$\oslash$n
\cite{Gron} and Maartens \cite{Maartens} presented exhaustive review
on cosmological models with non-causal and causal thermodynamics,
respectively. Later on, the cosmological models concerning the
particle production and bulk viscosity was studied by several
researchers \cite{Arbab, Maartensmend, Singhgp, Singhgpi, Bali,
Tyagi, Hubl, Lima, Zimdahl, Triginer, Johri}. Several other authors
have also explored the idea of the cosmological models with bulk
viscosity and particle creation. Singh and Kale \cite{Singhkale}
have studied the anisotropic bulk viscous cosmological models with
particle creation in general relativity. Further they have solved
the field equation of Brans-Dicke theory for Bianchi-type I space
model considering the bulk viscosity with particle creation in
energy momentum tensor of the cosmic fluid \cite{Singhkalei}.
Recently Chaubey \cite{Chaubey} has found the solution for Bianchi
type-V bulk viscous cosmological models with particle
creation in Brans-Dicke theory.\\

\noindent Now we study the particle creation and bulk viscosity for
early universe in details. Let us introduce the effective energy
momentum tensor $T_{ij}$ of the standard Einstein's field equation
in the presence of creation of particle and bulk viscosity, which
includes the creation pressure term $p_c$ and the bulk viscous
stress $\Pi$ as follows:
\begin{equation}
T_{\mu\nu} = \left(\rho +p+p_c + \Pi \right)u_{\mu}u_{\nu} - \left(p+p_c+\Pi\right) g_{\mu\nu}
\end{equation}
where $\rho$ is the energy-density, $p$ the pressure. Also the four velocity vector of the fluid following
 $u^{\mu}u_{\mu}=1$.\\

\noindent The particle number density flow vector $N^i$ $(=\eta u^i)$ and entropy flux vector $S^i$ $(=\nu \eta u^i)$ in second
 law of thermodynamics suggest the following equations respectively
\begin{equation}
N^{i}_{;i}=\dot{\eta}+3\eta H= \Gamma
\end{equation}
\begin{equation}
S^{i}_{;i}=\eta \dot{\nu}+\nu \Gamma \geq 0.
\end{equation}
Here $\eta$ is particle number density, $\nu$ is entropy per particle, $H$ is Hubble parameter and $\Gamma$ is the source term
 which will be positive if there is production of particles and it is negative when there is annihilation of particles.
  But it will vanish if there is no particle production or annihilation. A dot$(.)$ here denotes the differentiation
  with respect to time.    \\

\noindent For an open adiabatic system to cosmology, the supplementary pressure $p_c$, due to the creation of matter,
assumes the following form \cite{Prigoginei}
\begin{equation}
p_c=-\frac{(\rho+p)}{3H\eta}\Gamma= -\frac{(\rho+p)}{3H}\left(3H+\frac{\dot{\eta}}{\eta}\right).
\end{equation}
For an open thermodynamical system, the Gibbs equation\cite{Maartens} can be given by
\begin{equation}
\eta T\dot{\nu}=\dot{\rho}-(\rho+p)\frac{\dot{\eta}}{\eta}
\end{equation}
where $T$ is the temperature of the cosmic fluid. Using the continuity equation (23)[to be introduced in the next section],
 equation (2) and equation (5), the expression for entropy per particle is
\begin{equation}
\dot{\nu}=-\frac{3Hp_c}{\eta T}-\frac{3H\Pi}{\eta T}-\frac{(\rho+p)}{\eta^2 T}\Gamma.
\end{equation}
From equations (4) and (6), the entropy per particle can be calculated in more simplified form as
\begin{equation}
\dot{\nu}=-\frac{3H\Pi}{\eta T}.
\end{equation}
From equations (5) and (7), we have the following relation
\begin{equation}
\frac{\dot{\eta}}{\eta}=\frac{\dot{\rho}+3H\Pi}{\rho+p}.
\end{equation}
Using the equation of state $p=\gamma \rho(0\leq\gamma\leq1)$ in the above equation (8), the particle number density
can be obtained as
\begin{equation}
\eta^{1+\gamma}=\eta_0 \rho \exp{\left(\int 3H\Pi \rho^{-1}dt\right)}
\end{equation}
where $\eta_0$ is an integration constant. The conventional bulk
viscous effect in Bianchi universe can be modeled within the
framework of non-equilibrium thermodynamics. Here the transport
equation for the bulk viscous pressure $\Pi$ takes the
form\cite{Maartens}
\begin{equation}
\Pi+\tau \dot{\Pi}=-3\zeta H-\frac{\epsilon \tau \Pi}{2}\left(3H+\frac{\dot{\tau}}{\tau}
-\frac{\dot{\zeta}}{\zeta}-\frac{\dot{T}}{T}\right)
\end{equation}
where $\zeta$ is the bulk viscous coefficient and $\tau$ is the relaxation time associated with the dissipative effect.
 In view of the above relationship, we can study the behaviour of Bulk Viscosity in \textbf{Truncated theory},
 \textbf{Full Causal theory} and \textbf{Eckart's theory} by putting $\epsilon=0$, $\epsilon=1$ and $\tau=0$ respectively
  in the above equation. In truncated theory the above evolution equation (10) reduces to
\begin{equation}
\Pi+\tau \dot{\Pi}=-3\zeta H.
\end{equation}
Also in this theory, to ensure that the viscous signals do not exceed the speed of light, we consider the following relation
\begin{equation}
\tau=\frac{\zeta}{\rho}.
\end{equation}
In Full Causal theory, the equation of state for pressure and temperature \cite{Maartens, Maartensmend} are
taken to be barotropic  i.e., $p=\gamma \rho$ and $T=T(\rho)$. Then $T\propto exp \int \frac{dp(\rho)}{\rho+p(\rho)}$,
this will reduce to the following equation by using the equation of state for pressure as
\begin{equation}
T= T_0 \rho^{\frac{\gamma}{(1+\gamma)}}
\end{equation}
where $T_0$ is a constant. Using equations (12) and (13), the evolution equation (10) will take the form
\begin{equation}
\Pi+\frac{\zeta}{\rho} \dot{\Pi}= -3H \zeta-\frac{\zeta \Pi}{2\rho}\left[3H-\frac{(1+2\gamma)\dot{\rho}}{(1+\gamma)\rho}\right].
\end{equation}
In Eckart's non-causal theory, the evolution equation (10) will reduce to
\begin{equation}
\Pi=-3\zeta H.
\end{equation}
In the following sections, we solve the standard Einstein's field equation for an anisotropic Bianchi type V model in the
presence of particle creation and bulk viscosity.\\

\section{Basic Equations}
As a gravitational field we consider the Bianchi type V space-time given by
\begin{equation}
ds^2=dt^2-A^{2}dx^2-e^{2m x}[B^{2}dy^2+C^{2}dz^2].
\end{equation}
The Einstein's field equation can be written as
\begin{equation}
R_{\mu\nu}-\frac{1}{2}g _{\mu\nu}R =-8\pi G T_{\mu\nu}
\end{equation}
all the symbols above have their usual meaning. The Einstein's field equation (17) in Bianchi V line element (16)
 for the energy-momentum tensor (1) can be translated into the following set of non-linear equations as
\begin{eqnarray}
\frac{\ddot{B}}{B}+\frac{\ddot{C}}{C}+\frac{\dot{B}}{B}\frac{\dot{C}}{C}-\frac{m^2}{A^{2}}=-8\pi G ( p+p_c+\Pi)
\end{eqnarray}

\begin{eqnarray}
\frac{\ddot{A}}{A}+\frac{\ddot{C}}{C}+\frac{\dot{A}}{A}\frac{\dot{C}}{C}-\frac{m^2}{A^{2}}=-8\pi G (p+p_c+\Pi)
\end{eqnarray}

\begin{eqnarray}
\frac{\ddot{A}}{A}+\frac{\ddot{B}}{B}+\frac{\dot{A}}{A}\frac{\dot{B}}{B}-\frac{m^{2}}{A^{2}}=-8\pi G (p+p_c+\Pi)
\end{eqnarray}

\begin{eqnarray}
\frac{\dot{A}}{A}\frac{\dot{B}}{B}+\frac{\dot{A}}{A}\frac{\dot{C}}{C}+\frac{\dot{B}}{B}\frac{\dot{C}}{C}
-\frac{3m^{2}}{A^{2}}
= 8\pi G \rho
\end{eqnarray}

\begin{eqnarray}
\frac{\dot{B}}{B}+\frac{\dot{C}}{C}-2\frac{\dot{A}}{A}=0.
\end{eqnarray}
The Bianchi identity reads
\begin{eqnarray}
\dot{\rho}+3(\rho+p)H= -3(p_{c}+\Pi)H.
\end{eqnarray}
Here we consider that Hubble parameter $H$ is directly proportional to a negative n power of average scale factor $a$ so that
 we have a relationship between these parameters given by $H=h_0 a^{-n}=h_0 (ABC)^{-n/3}$ where $h_0>0$ and $n\geq 0$ are constants.
  This type of relation gives a constant value of deceleration parameter which is a very useful tool in solving the field equations.
   Earlier Berman \cite{Berman}, Berman and Gomide \cite{Bermangomide} had considered such type of assumption for
   solving FRW cosmological models. Later on, many workers \cite{Singhcp, Shriram, Shrirami, Shriramij} have used this
    assumption for solving Einstein's field equations in general relativity and different scalar tensor theory of gravitation.
    The general formulas of certain physical parameters for the metric equation (16) are given as follows:\\

\noindent The expansion scalar is given by

\begin{equation}
\theta = u ^{\mu}_{;\mu}=\frac{\dot{A}}{A}+\frac{\dot{B}}{B}+\frac{\dot{C}}{C}.
\end{equation}
The shear scalar has the form

\begin{equation}
\sigma ^{2} = \frac {1}{2}\sigma _{\mu\nu}\sigma^{\mu\nu}= \frac {1}{2}\left[\left(\frac{\dot{A}}{A}\right)^{2}
+  \left(\frac{\dot{B}}{B}\right)^{2}+\left(\frac{\dot{C}}{C}\right)^{2}\right]-\frac{{\theta}^{2}}{6}.
\end{equation}
We also introduce generalized Hubble parameter $H$:
\begin{equation}
H= \frac{1}{3}\left(H_{1}+H_{2}+H_{3}\right)
\end{equation}
with $H_{1}=\frac{\dot{A}}{A}$, $H_{2}=\frac{\dot{B}}{B}$ and
$H_{3}=\frac{\dot{C}}{C}$  are the directional Hubble parameters  in the
directions of $x$, $y$ and $z$ respectively. Let us introduce the
function $V$ as
\begin{equation}
V=ABC.
\end{equation}
The average scale factor $a$ can be written in terms of metric functions as
\begin{equation}
a=(ABC)^{1/3} = V^{1/3}.
\end{equation}
It should be noted that the parameters $H$, $V$ and $a$ are connected by the following relation

\begin{equation}
H=\frac{1}{3}\frac{\dot{V}}{V}=\frac{\dot{a}}{a}.
\end{equation}
The deceleration parameter $q$ is given as
\begin{equation}
q=-\frac{a\ddot{a}}{{\dot{a}}^2}.
\end{equation}
Using the relation $H=h_0 a^{-n}$ and the above equations, the expressions for deceleration parameter $q$ and average
scale factor $a$ can be calculated as
\begin{equation}
q=n-1.
\end{equation}
\begin{equation}
a=(nh_0t)^{1/n}, n\neq 0
\end{equation}
and
\begin{equation}
a=a_0 \exp(h_0t), n=0
\end{equation}
where $a_0$ is an integration constant. Now we follow the approach of Saha and Rikhvitsky \cite{Saha} and Shri Ram et al.
 \cite{Shriramij}. Hence from equations (18)-(22), the quadrature form of the metric functions can be given as follows:
\begin{equation}
A(t)=a
\end{equation}
\begin{equation}
B(t)=B_0 a  \exp\left(\frac{X}{3}\int{\frac{dt}{a^3}}\right)
\end{equation}
\begin{equation}
C(t)=C_0 a  \exp\left(-\frac{X}{3}\int{\frac{dt}{a^3}}\right)
\end{equation}
where $B_0$, $C_0$ and $X$ are arbitrary constants. Taking into
account that
$$BC = A^2$$
which is the consequence of the equation(22) [for simplicity we set
the integration constant to be unit], we find $B_0 C_0 = 1$.

Now in the following sections we present the exact solutions of the
above field equations in view of two different values of average
scale factor given by equations (32) and (33).
 In this way, we can formulate two basic cosmological models. These models can also be studied in context with different
  physical behaviors of the particle creation and bulk viscosity in the energy momentum tensor of the cosmic fluid.
   That will create various sub-models.
\section{The Cosmological Models}
\subsection{Model 1}
Putting the value of the average scale-factor from equation (32) into equations (34)-(36), the exact values of
 the metric functions can be obtained as
\begin{equation}
A=(nh_0t)^{1/n}
\end{equation}
\begin{equation}
B=B_0 (nh_0t)^{1/n} \exp{\left[\frac{X}{3h_0(n-3)}(nh_0t)^{\frac{n-3}{n}}\right]}
\end{equation}
\begin{equation}
C=C_0 (nh_0t)^{1/n} \exp{\left[-\frac{X}{3h_0(n-3)}(nh_0t)^{\frac{n-3}{n}}\right]}
\end{equation}
where $n\neq 3$.

\myfigures{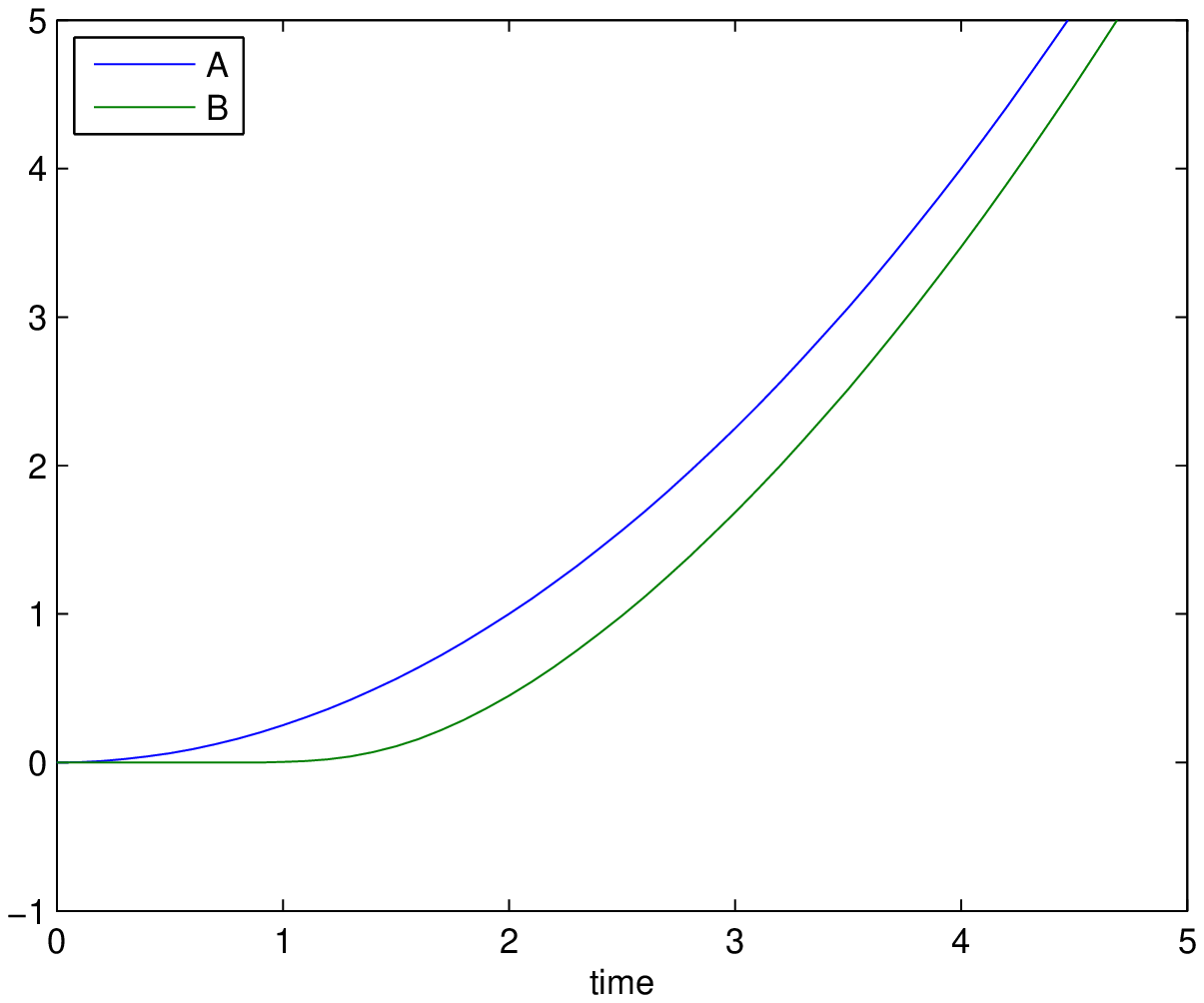}{0.45}{Variation of scale factors $A$ and $B$ for the
parameter $n=0.5$ with time.} {0.45}{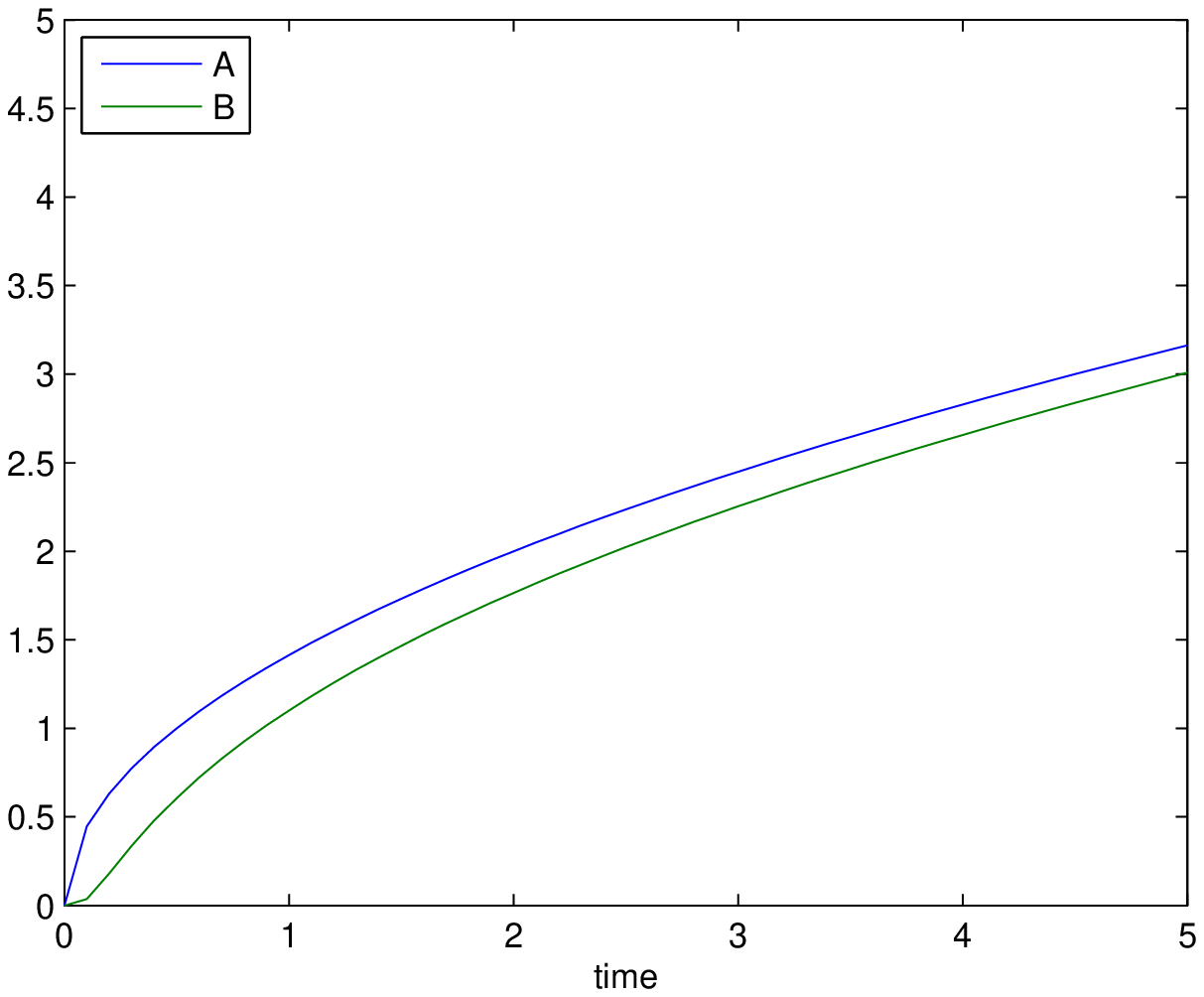}{0.45}{Variation of scale
factors $A$ and $B$ for the parameter $n=2$ with time.}{0.45}

\noindent Hence the Expansion scalar $\theta$, Shear scalar
$\sigma^2$, generalized Hubble parameter $H$ and the Volume scalar
$V$ can be written as
\begin{equation}
\theta=3h_0(nh_0t)^{-1}
\end{equation}
\begin{equation}
\sigma^2=\frac{X^2}{9}(nh_0t)^{-6/n}
\end{equation}
\begin{equation}
H=h_0(nh_0t)^{-1}
\end{equation}
and
\begin{equation}
V=(nh_0t)^{3/n}.
\end{equation}
The anisotropy parameter $A_m$ can be given by
\begin{equation}
A_m=\frac{2}{27}\frac{X^2}{h_0^{2}}(nh_0t)^{\frac{2n-6}{n}}.
\end{equation}
The directional Hubble parameters can be obtained as
\begin{equation}
H_1=h_0(nh_0t)^{-1}
\end{equation}
\begin{equation}
H_2=h_0(nh_0t)^{-1}+\frac{X}{3}(nh_0t)^{-3/n}
\end{equation}
\begin{equation}
H_3=h_0(nh_0t)^{-1}-\frac{X}{3}(nh_0t)^{-3/n}.
\end{equation}
From equation(21), the value of the energy density can be found as
\begin{equation}
\rho=\rho_0(nh_0t)^{-2}-\rho_1(nh_0t)^{-6/n}-\rho_2(nh_0t)^{-2/n}
\end{equation}
where $\rho_0=\frac{3h_0^{2}}{8\pi G}$, $\rho_1=\frac{X^2}{72\pi G}$, $\rho_2=\frac{3m^2}{8\pi G}$.
 Due to the barotropic equation $p=\gamma \rho$, we have the following expression for pressure as
\begin{equation}
p=\gamma\left[\rho_0(nh_0t)^{-2}-\rho_1(nh_0t)^{-6/n}-\rho_2(nh_0t)^{-2/n}\right].
\end{equation}

\noindent We now investigate the behavior of the above cosmological
model by analyzing the different physical parameters. The behaviours
of the scale factors $A$, $B$ and $C$ can be observed in the figures
1, 2 and 3. The scale factors $A$ and $B$ are increasing function of
time. They are accelerated and decelerated for the parameter
 $n=0.5$ and $n=2$ respectively. Above results show that all three scale factors
  are zero at the initial time $t=0$. Expansion scalar, Shear scalar, Hubble parameter and the three directional Hubble
  parameters are all infinite at $t=0$. It is also observed that the spatial volume is zero at this initial time.
  The mean anisotropy parameter also diverge at this time for $n<3$. The energy density and pressure tend to infinity at this epoch.
   All these values of different physical parameters show that the universe starts evolving with zero volume and expands with
   cosmic time $t$. That is the model has point singularity at $t=0$. Now we also study these parameters for very large time
   as $t \rightarrow \infty$. We find here that $A$ and $B$ tend to zero but $C$ will become indeterminate. Expansion scalar,
    Shear scalar, Hubble parameter, the three directional Hubble parameters and the mean anisotropy parameter will all become
    zero for large time. The energy density and pressure tend to zero at $t \rightarrow \infty$. All these indicate that
    the universe is expanding with the increase of cosmic time but the rate of expansion and shear scalar decrease to zero
    and finally tend to isotropic. This model approaches isotropic during late time of its evolution
     as $\lim \sigma^{2}/\theta= 0$ for $t\rightarrow \infty$.\\

\noindent Now in the following subsections we study the behavior of
particle creation and bulk viscosity of this model in four different
physical laws. Therefore, we can create four different sub models
out of the above model.
\subsubsection{Bulk Viscosity Energy-Density Law}
Let us assume here that the bulk viscous stress $\Pi$ be associated with energy density $\rho$ by the following relationship
\begin{equation}
\Pi=\Pi_0 \rho^{\omega}
\end{equation}
where $\Pi_0$ is a constant. The above assumption is motivated by the relation $\zeta=\zeta_0 \rho^{\omega}$
where $\zeta_0\geq 0, \omega\geq 0$. This expression for $\zeta$ was suggested by several researchers
\cite{Johrisudharsan, Maartensr, Pavon}. Further they added that for $\omega=1$, this will correspond
to radiative fluid and for $\omega=1.5$, this may represent a string dominated universe \cite{Murphy, Santosh}.
 So, here the expression for $\Pi$ can be written as
\begin{equation}
\Pi=\Pi_0 \left[\rho_0(nh_0t)^{-2}-\rho_1(nh_0t)^{-6/n}-\rho_2(nh_0t)^{-2/n}\right]^{\omega}.
\end{equation}
The graphical behaviour of the above equation for $\Pi$ can be
observed through the figure 4 in different parameters. The creation
pressure can be obtained as
\begin{eqnarray}
p_c &=& p_0(nh_0t)^{-2}+p_1(nh_0t)^{-2/n}+p_2 (nh_0t)^{-6/n}\nonumber \\
&& - \Pi_0 \left[\rho_0(nh_0t)^{-2}-\rho_1(nh_0t)^{-6/n}-\rho_2(nh_0t)^{-2/n}\right]^{\omega}
\end{eqnarray}
where $p_0= \left[\frac{2n}{3}-(1+\gamma)\right]\rho_0$, $p_1=\frac{(3\gamma+1)\rho_2}{3}$ and $p_2=(\gamma-1)\rho_1$.
The variation of creation pressure in this case can be observed in figures from 5 to 10.\\

\myfigures{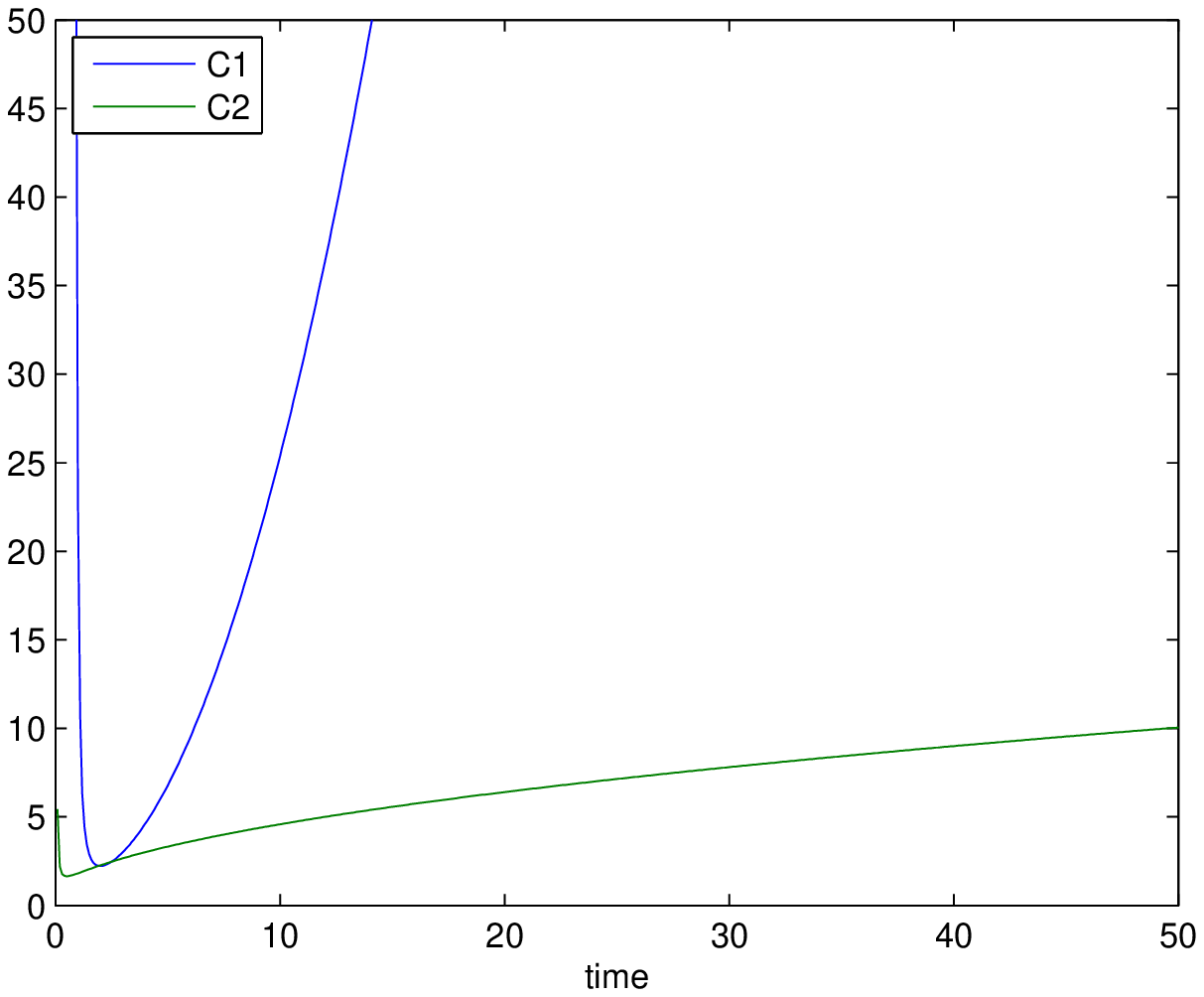}{0.45}{Variation of the scale factor $C$ for $n=0.5$
and $n=2$ with time.} {0.45}{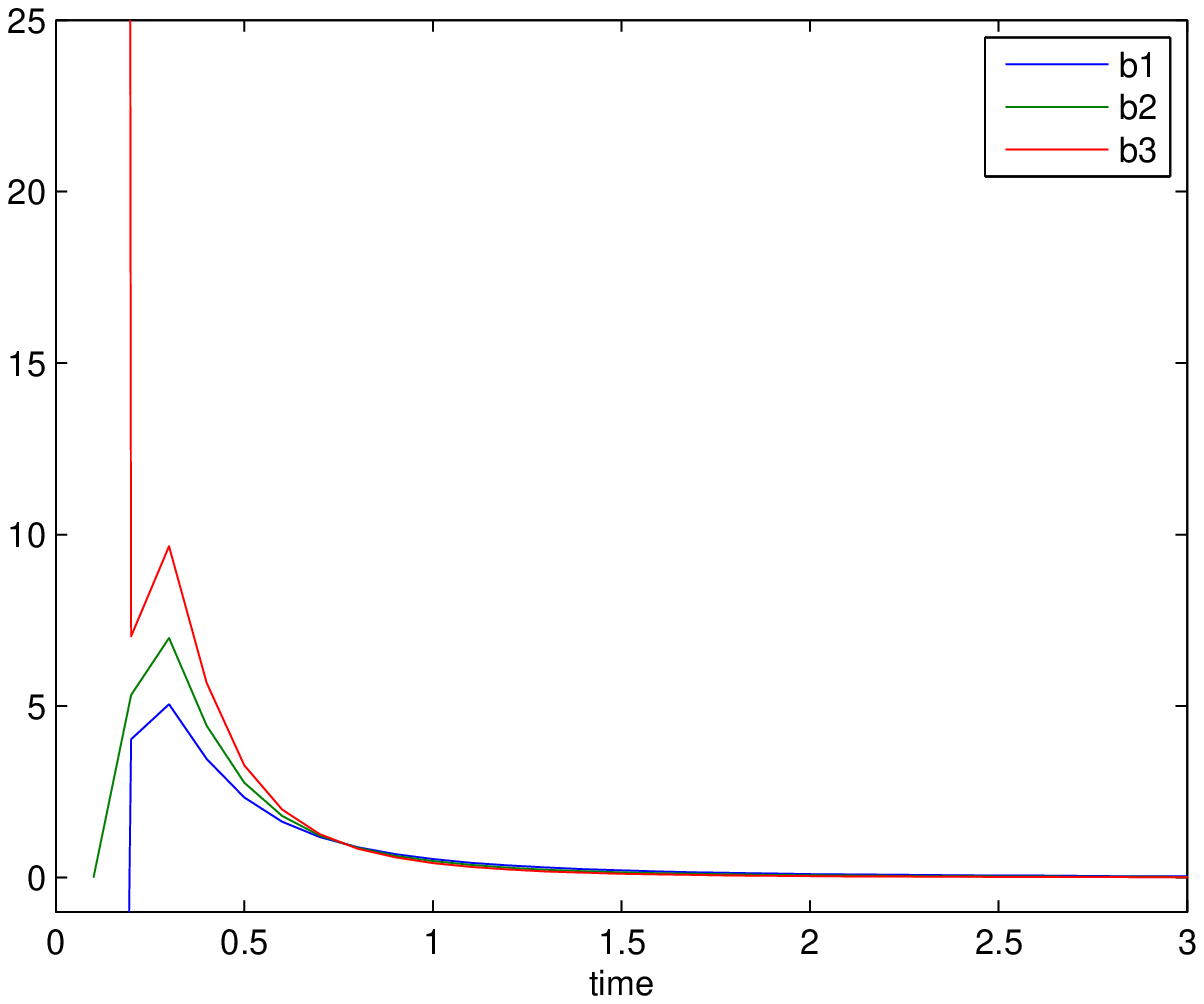}{0.45}{Variation of bulk viscous
stress $\Pi$ with time for the subcase Bulk Viscosity Energy-Density
Law in Model 1. $b1$, $b2$ and $b3$ represent variation for the
parameter $\omega=1.25$, $\omega=1.5$ and $\omega=1.75$
respectively.}{0.45}

\myfigures{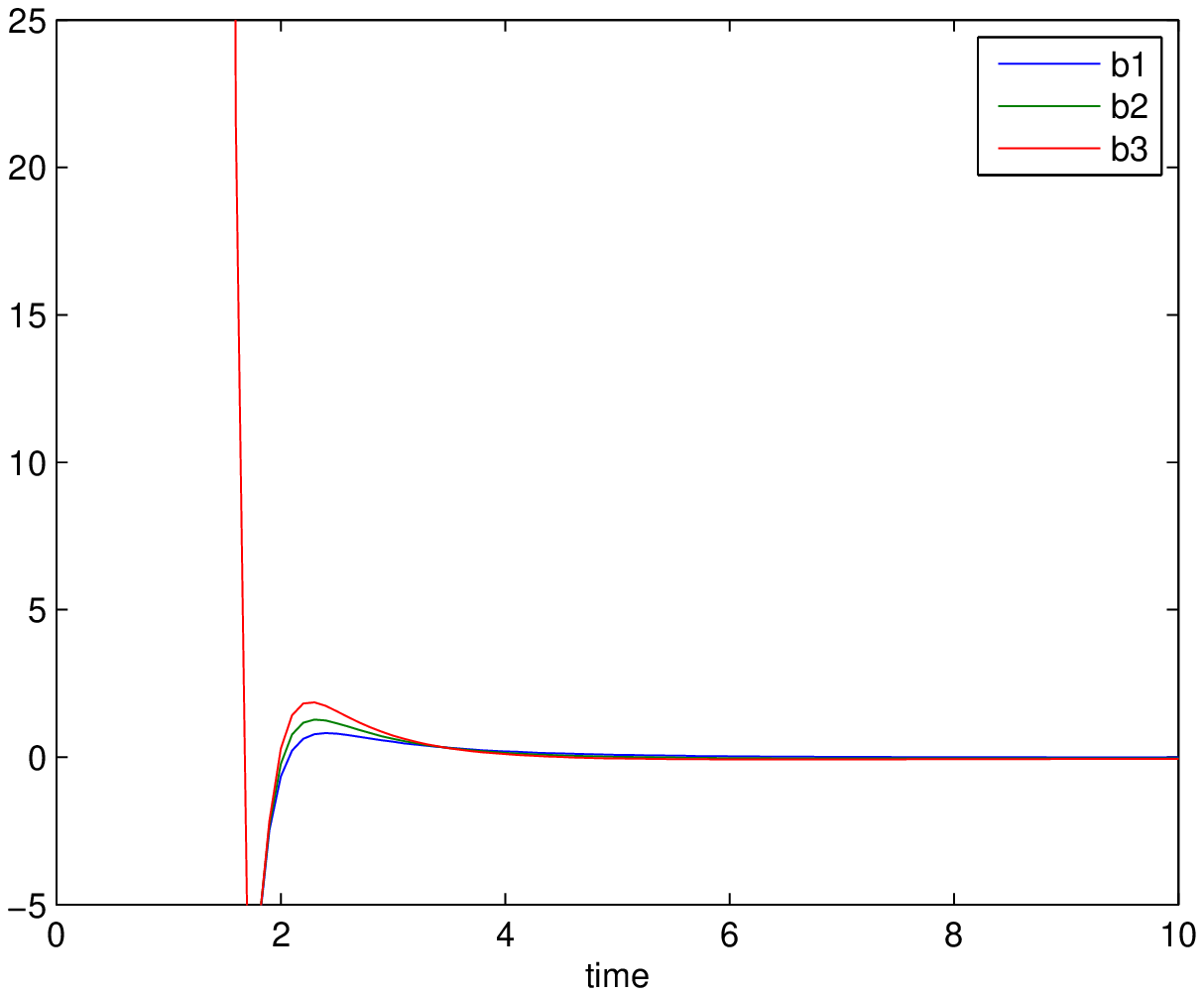}{0.45}{Variation of creation pressure $Pc$ for
$n=0.5$ and $\gamma=0$ with time for the subcase Bulk Viscosity
Energy-Density Law in Model 1. $b1$, $b2$ and $b3$ represent
variation for the parameter $\omega=1.25$,
 $\omega=1.5$ and $\omega=1.75$ respectively.}
{0.45}{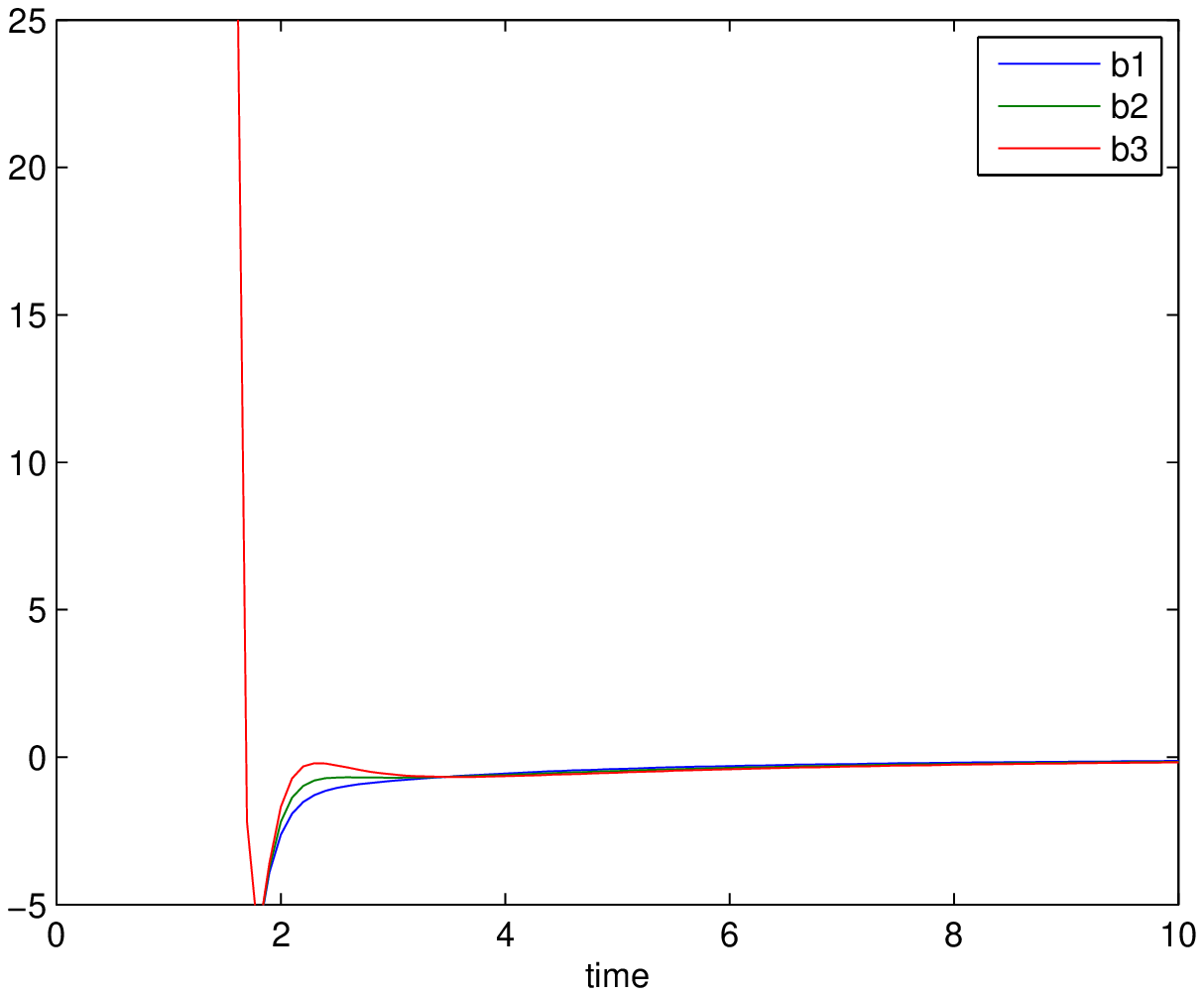}{0.45}{Variation of creation pressure $Pc$ for
$n=0.5$ and $\gamma=1$ with time for the subcase Bulk Viscosity
Energy-Density Law in Model 1. $b1$, $b2$ and $b3$ represent
variation for the parameter $\omega=1.25$, $\omega=1.5$ and
$\omega=1.75$ respectively.}{0.45}

\myfigures{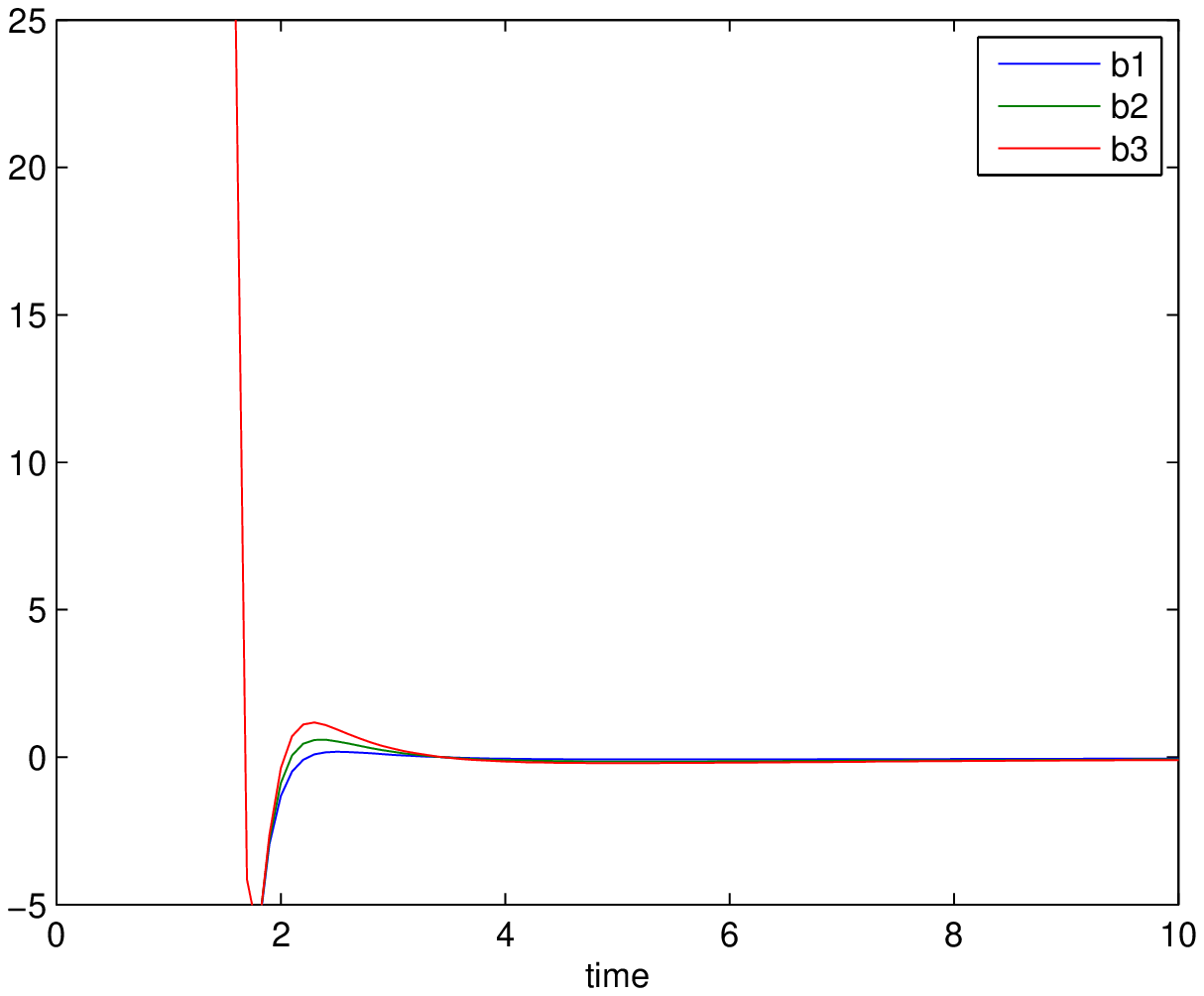}{0.45}{Variation of creation pressure $Pc$
for $n=0.5$ and $\gamma=1/3$ with time for the subcase Bulk
Viscosity Energy-Density Law in Model 1. $b1$, $b2$ and $b3$
represent variation for the parameter $\omega=1.25$, $\omega=1.5$
and $\omega=1.75$ respectively.} {0.45}{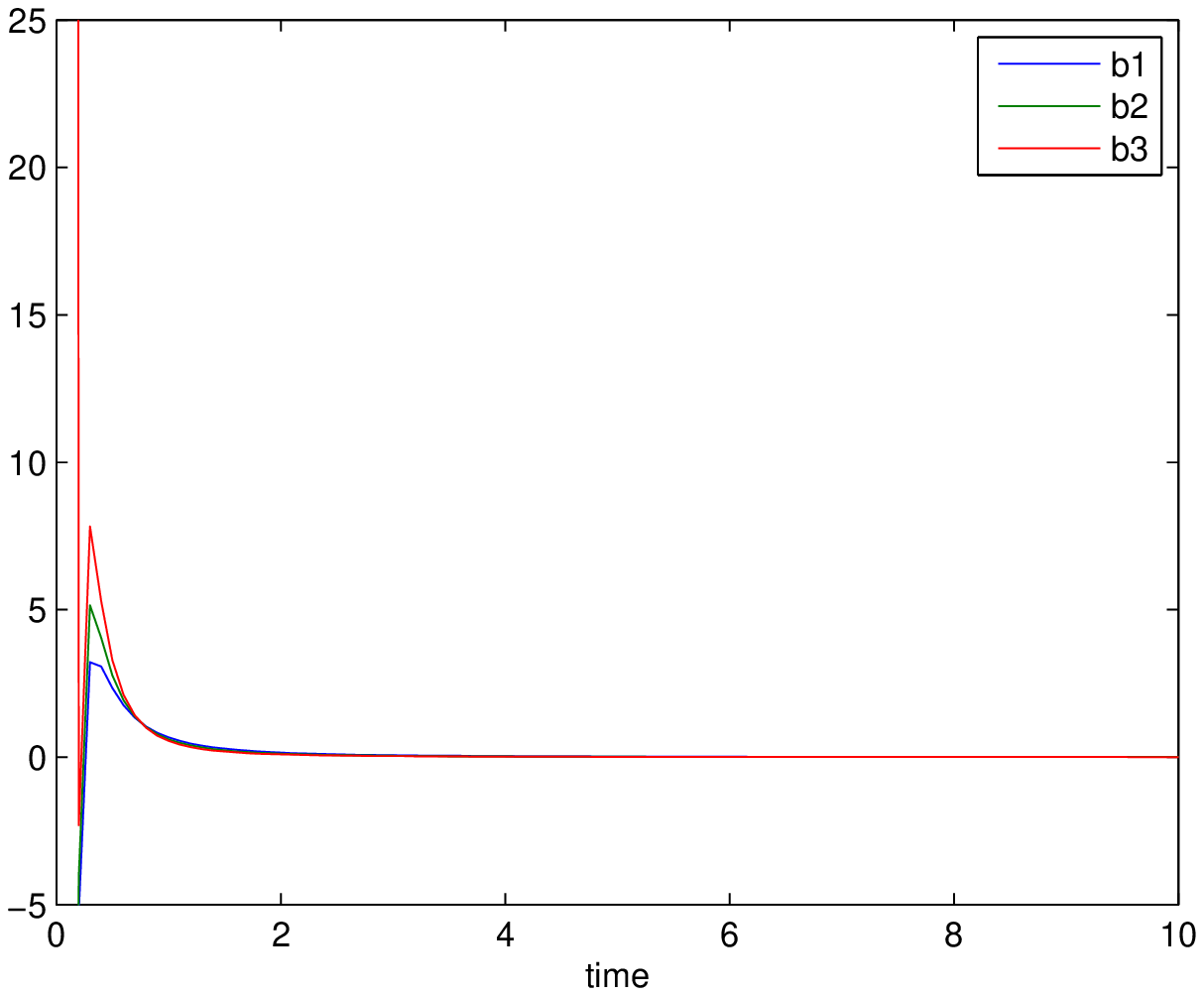}{0.45}{Variation of
creation pressure $Pc$ for $n=2$ and $\gamma=0$ with time for the
subcase Bulk Viscosity
 Energy-Density Law in Model 1. $b1$, $b2$ and $b3$ represent variation for the parameter $\omega=1.25$,
$\omega=1.5$ and $\omega=1.75$ respectively.}{0.45}

\myfigures{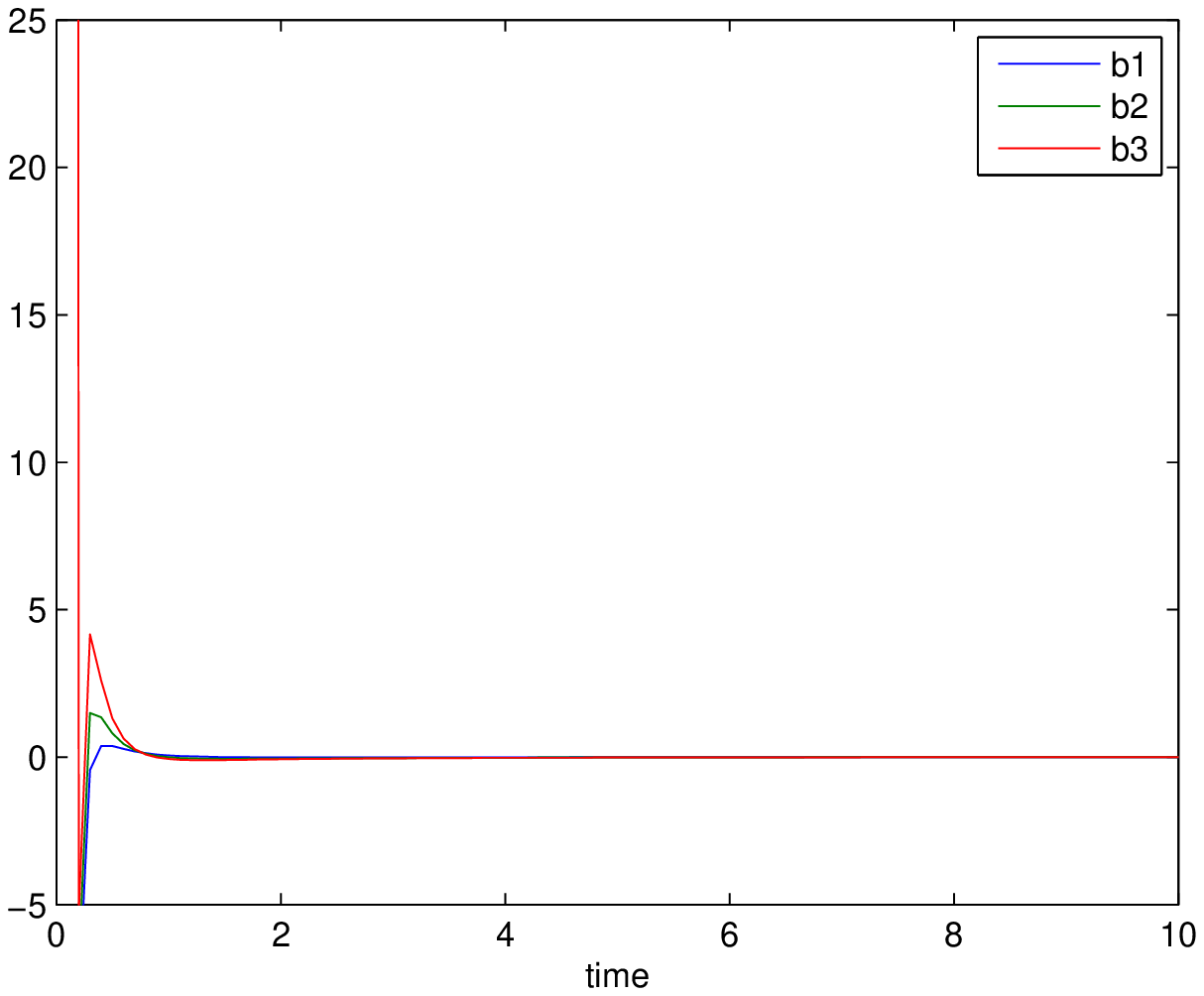}{0.45}{Variation of creation pressure $Pc$ for
$n=2$ and $\gamma=1$ with time for the subcase Bulk Viscosity
Energy-Density Law in Model 1. $b1$, $b2$ and $b3$ represent
variation for the parameter $\omega=1.25$, $\omega=1.5$ and
$\omega=1.75$ respectively.}{0.45}{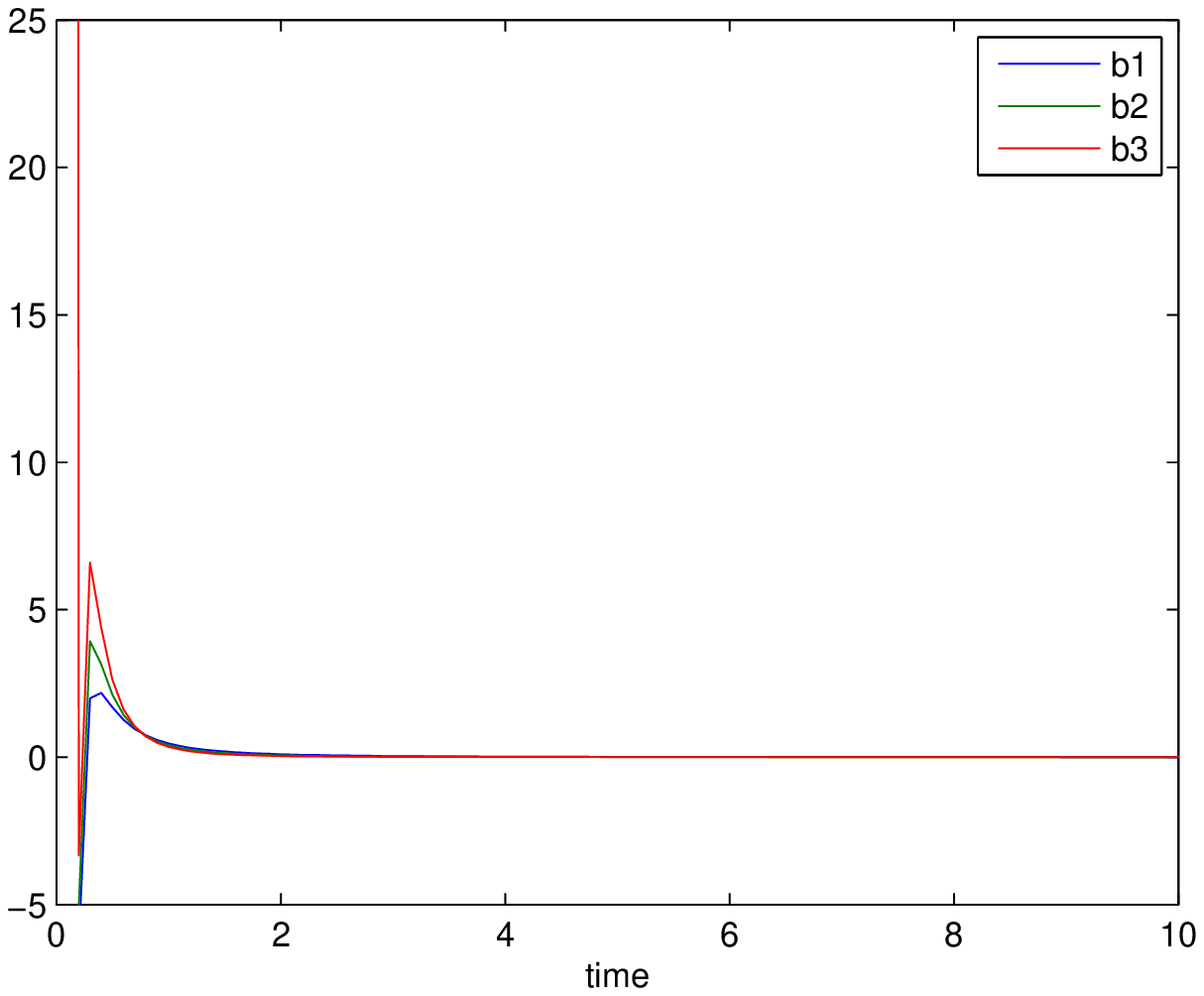}{0.45}{Variation of
creation pressure $Pc$ for $n=2$ and $\gamma=1/3$ with time for the
subcase Bulk Viscosity Energy-Density Law in Model 1. $b1$, $b2$ and
$b3$ represent variation for the parameter $\omega=1.25$,
$\omega=1.5$ and $\omega=1.75$ respectively.}{0.45}

\noindent The value of bulk viscosity coefficient in Truncated theory can be written as
\begin{eqnarray}
\zeta &=& \zeta_1\left[\rho_0(nh_0t)^{-3}-\rho_1(nh_0t)^{\frac{-6}{n}-1}-\rho_2 (nh_0t)^{\frac{-2}{n}-1}\right]\nonumber \\
&& +\zeta_2 \left[\rho_0(nh_0t)^{-2}-\rho_1(nh_0t)^{\frac{-6}{n}}-\rho_2(nh_0t)^{\frac{-2}{n}}\right]^{\omega-1}\nonumber \\
&&
\left[\frac{3\rho_1}{n}(nh_0t)^{\frac{-6}{n}-1} +\frac{\rho_2}{n}(nh_0t)^{\frac{-2}{n}-1}-\rho_0 (nh_0t)^{-3}\right].
\end{eqnarray}
Here $\zeta_1=3 h_0$ and $\zeta_2= 2nh_0 \omega \Pi_0$.\\

\noindent The bulk viscosity coefficient in Full Causal theory can be obtained as
\begin{equation}
\zeta= \frac{\displaystyle \zeta_3\left[\rho_0(nh_0t)^{-2}-\rho_1(nh_0t)^{\frac{-6}{n}}
-\rho_2(nh_0t)^{\frac{-2}{n}}\right]^{\omega+2}  }{\displaystyle
 \begin{array}{c}6h_0(nh_0t)^{-1}\left[\rho_0(nh_0t)^{-2}-\rho_1(nh_0t)^{\frac{-6}{n}}
 -\rho_2 (nh_0t)^{\frac{-2}{n}}\right]^{2} \\
 +\zeta_4\left[\rho_0(nh_0t)^{-3}-\frac{3\rho_1}{n}(nh_0t)^{\frac{-6}{n}-1}-\frac{\rho_2}{n}
 (nh_0t)^{\frac{-2}{n}-1}\right]\\
 +\zeta_5(nh_0t)^{-1}\left[\rho_0(nh_0t)^{-2}-\rho_1(nh_0t)^{\frac{-6}{n}}-\rho_2(nh_0t)^{\frac{-2}{n}}\right]^{\omega+1}\\
 +\zeta_6\left[\rho_0(nh_0t)^{-2}-\rho_1(nh_0t)^{\frac{-6}{n}}-\rho_2(nh_0t)^{\frac{-2}{n}}\right]^{\omega-1}\\
 \left[-\rho_0(nh_0t)^{-3}+\frac{3\rho_1}{n}(nh_0t)^{\frac{-6}{n}-1}+\frac{\rho_2}{n}
 (nh_0t)^{\frac{-2}{n}-1}\right]\\
 \end{array}}
\end{equation}
where $\zeta_3= -2\Pi_0$, $\zeta_4=\frac{2nh_0(1+2\gamma)}{(1+\gamma)}$, $\zeta_5=3\Pi_0h_0$ and $\zeta_6=4n h_0 \Pi_0 \omega$. \\

\noindent Similarly the bulk viscosity coefficient in Eckart theory can be given as
\begin{eqnarray}
\zeta=\frac{-\Pi_0\left[\rho_0(nh_0t)^{-2}-\rho_1(nh_0t)^{\frac{-6}{n}}-\rho_2(nh_0t)^{\frac{-2}{n}}\right]^{\omega}}
{3h_0(nh_0t)^{-1}}.
\end{eqnarray}
\subsubsection{Uniform Particle Number Density $(\dot{\eta}=0)$}
In this case of study we consider the particle number density to be uniform during the evolution of the universe,
 i.e., $(\dot{\eta}=0)$. This assumption gives the following values of the particle production term $\Gamma$ and the
 creation pressure $p_c$ as
\begin{equation}
\Gamma=3H\eta
\end{equation}
\begin{equation}
p_c=-(1+\gamma)\rho.
\end{equation}
So here the value of bulk viscous stress and the creation pressure can be obtained as
\begin{equation}
\Pi=\frac{2\rho_0 n}{3}(nh_0t)^{-2}-2\rho_1(nh_0t)^{-6/n}-\frac{2\rho_2}{3}(nh_0t)^{-2/n}.
\end{equation}
\begin{equation}
p_c=-(1+\gamma)\left[\rho_0(nh_0t)^{-2}-\rho_1(nh_0t)^{-6/n}-\rho_2(nh_0t)^{-2/n}\right].
\end{equation}

\myfigures{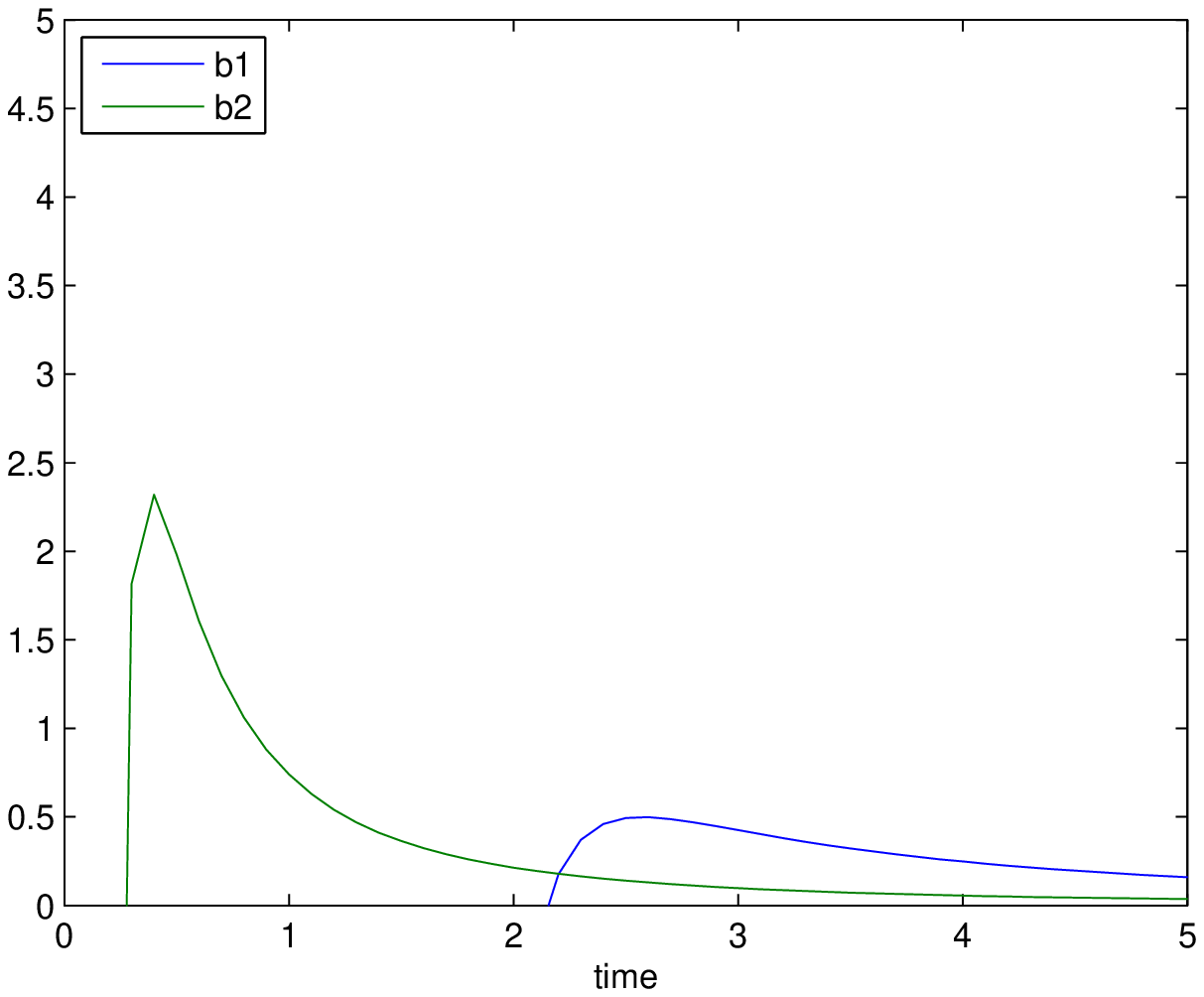}{0.45}{Variation of $\Pi$ with time for the subcase
Uniform Particle Number Density in Model 1. $b1$,
 $b2$ represent variation for the parameter $n=0.5$ and $n=2$.} {0.45}{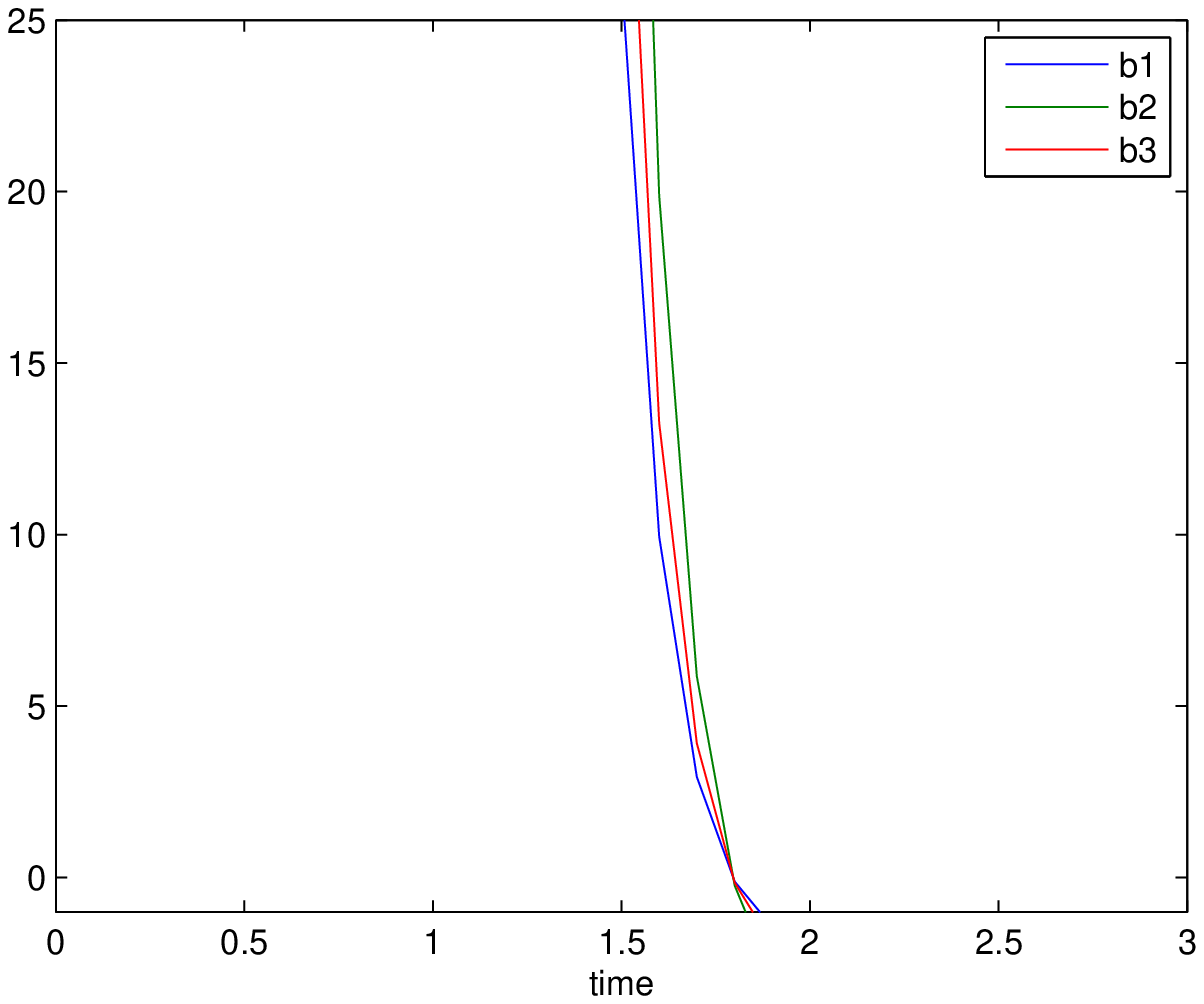}{0.45}{Variation of creation pressure
 $Pc$ for $n=0.5$ with time for the subcase Uniform Particle Number
 Density in Model 1. $b1$, $b2$ and $b3$ represent variation for the parameter $\gamma=0$, $\gamma=1$ and $\gamma=1/3$
 respectively.}{0.45}

\noindent The behavior of $\Pi$ and $Pc$ in Uniform Particle Number
Density for Model 1 can be seen in figures from 11 to 13.
The expression for bulk viscous coefficient in all three cases are as follows:\\

\noindent Truncated theory:\\
\begin{equation}
\zeta= \frac{\displaystyle
\begin{array}{c}\left[2\rho_1(nh_0t)^{\frac{-6}{n}}+\frac{2\rho_2}{3}
 (nh_0t)^{\frac{-2}{n}}-\frac{2\rho_0 n}{3}(nh_0t)^{-2}\right]\\
 \left[\rho_0(nh_0t)^{-2}-\rho_1(nh_0t)^{\frac{-6}{n}}-\rho_2(nh_0t)^{\frac{-2}{n}}\right]\\
 \end{array}}{\displaystyle
 \begin{array}{c} \zeta_7(nh_0t)^{-3}+\zeta_8(nh_0t)^{\frac{-6}{n}}-\zeta_9 (nh_0t)^{\frac{-2}{n}}\\
 \end{array}}.
\end{equation}
Here $\zeta_7= \left(3 h_0 \rho_0-\frac{4h_0\rho_0n^2}{3}\right)$, $\zeta_8 = 9\rho_1 h_0$ and $\zeta_9= \frac{5}{3}\rho_2 h_0$. \\

\noindent Full Causal theory:\\
\begin{equation}
\zeta= \frac{\displaystyle
\begin{array}{c}\left[2\rho_1(nh_0t)^{\frac{-6}{n}}+\frac{2\rho_2}{3}
 (nh_0t)^{\frac{-2}{n}}-\frac{2\rho_0 n}{3}(nh_0t)^{-2}\right]\\
 \left[\rho_0(nh_0t)^{-2}-\rho_1(nh_0t)^{\frac{-6}{n}}-\rho_2(nh_0t)^{\frac{-2}{n}}\right]^{2}\\
 \end{array}}{\displaystyle
 \begin{array}{c}\left[\frac{2\rho_0 n}{3}(nh_0t)^{-2}-2\rho_1(nh_0t)^{\frac{-6}{n}}-\frac{2\rho_2}{3}
 (nh_0t)^{\frac{-2}{n}}\right]\\
 \left[\zeta_{10}(nh_0t)^{-3}-\zeta_{11} (nh_0t)^{\frac{-2}{n}-1}-\zeta_{12}(nh_0t)^{\frac{-6}{n}-1}\right] \\
 +\left[-\zeta_{13}(nh_0t)^{-3}+\zeta_{14}(nh_0t)^{\frac{-6}{n}-1}+\zeta_{15} (nh_0t)^{\frac{-2}{n}-1}\right]\\
 \left[\rho_0(nh_0t)^{-2}-\rho_1(nh_0t)^{-6/n}-\rho_2(nh_0t)^{-2/n}\right]\\
 +3h_0 (nh_0t)^{-1}\left[\rho_0(nh_0t)^{-2}-\rho_1(nh_0t)^{-6/n}-\rho_2(nh_0t)^{-2/n}\right]^{2}\\
 \end{array}}
\end{equation}
where $\zeta_{10}= \frac{3}{2}\rho_0 h_0+\frac{n(1+2\gamma)}{(1+\gamma)}\rho_0 h_0$,
$\zeta_{11}= \frac{3}{2}h_0\rho_2+ \frac{(1+2\gamma)}{(1+\gamma)}h_0\rho_2$,
 $\zeta_{12}= \frac{3}{2}h_0\rho_1+ \frac{3(1+2\gamma)}{(1+\gamma)}h_0\rho_1$,
$\zeta_{13}=\frac{4\rho_0 n^2h_0}{3}$, $\zeta_{14}=12\rho_1 h_0$ and $\zeta_{15}= \frac{4h_0 \rho_2}{3}$.\\

\noindent Eckart theory:\\
\begin{equation}
\zeta= \frac{2\rho_1}{3h_0}(nh_0t)^{\frac{-6}{n}+1}+\frac{2\rho_2}{9h_0}(nh_0t)^{\frac{-2}{n}+1}
-\frac{2\rho_0 n}{9h_0}(nh_0t)^{-1}.
\end{equation}
\subsubsection{Ideal Gas}
The conservation of total particle number in standard cosmology can be given as
\begin{equation}
N^{i}_{;i}=\dot{\eta}+3\eta H= 0
\end{equation}
which leads to $\Gamma=0$ and $p_c=0$. Hence this model will only have bulk viscosity.
 The expression for particle number density and bulk viscous stress can be obtained as
\begin{equation}
\eta=\eta_1 t^{\frac{-3}{n}}
\end{equation}
and
\begin{equation}
\Pi=\zeta_{16} (nh_0 t)^{-2}+\zeta_{17} (nh_0t)^{-6/n}+\zeta_{18} (nh_0 t)^{-2/n}
\end{equation}
where $\eta_1$ is an integration constant. Here
$\zeta_{16}=\left[\frac{2n}{3}-(1+\gamma)\right]\rho_0$,
$\zeta_{17}=(\gamma-1)\rho_1$ and
$\zeta_{18}=\frac{(1+3\gamma)}{3}\rho_2$. The figures from 14 to 16
show the graphical behaviour of $\Pi$ for Ideal Gas in Model 1.

\myfigures{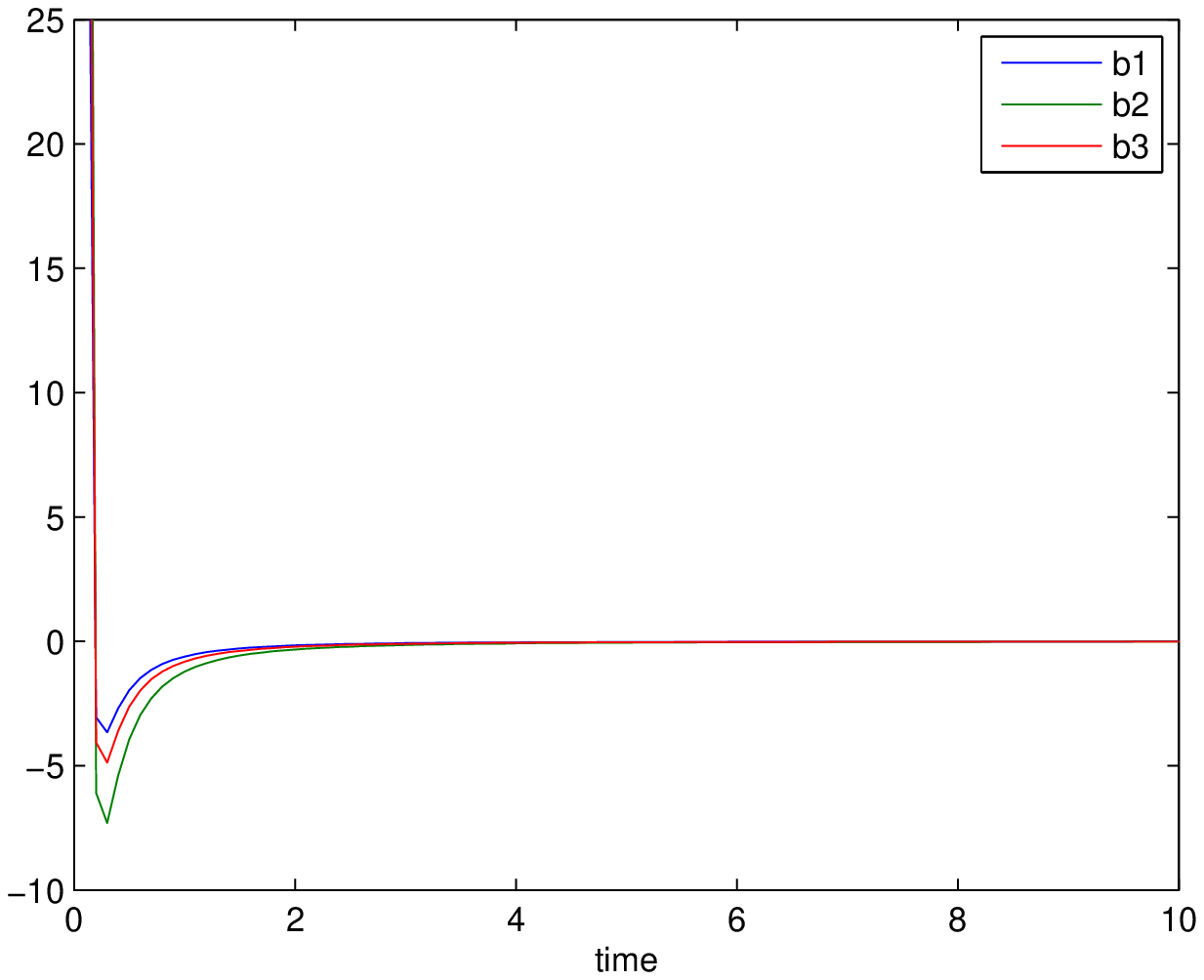}{0.45}{Variation of creation pressure $Pc$ for
$n=2$ with time for the subcase Uniform Particle Number Density in
Model 1. $b1$, $b2$ and $b3$ represent variation for the parameter
$\gamma=0$, $\gamma=1$ and $\gamma=1/3$ respectively.}
{0.45}{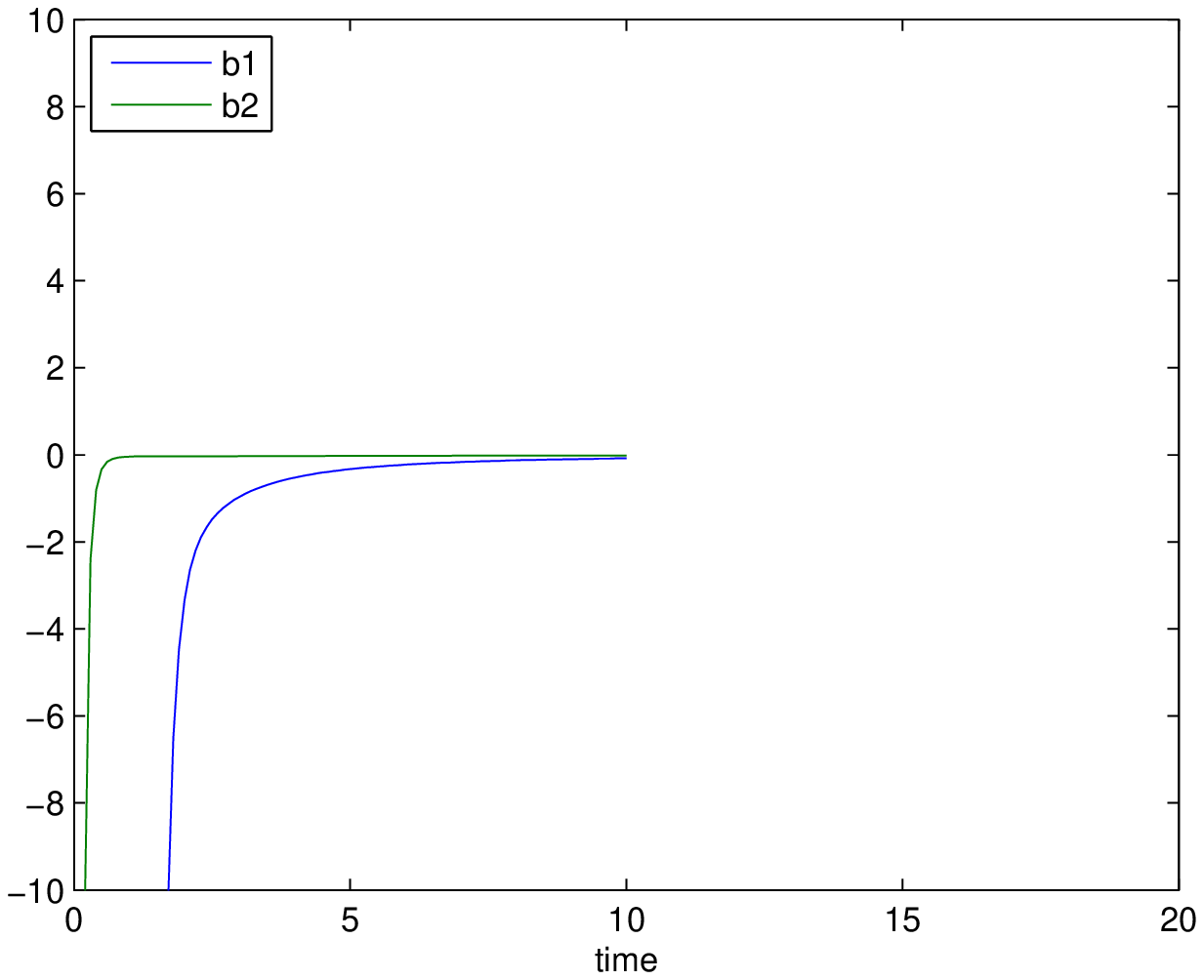}{0.45}{Variation of $\Pi$ for $\gamma=0$ with time for
the subcase Ideal Gas in Model 1. $b1$ and $b2$ represent variation
for the parameter $n=0.5$ and $n=2$ respectively.}{0.45}

\myfigures{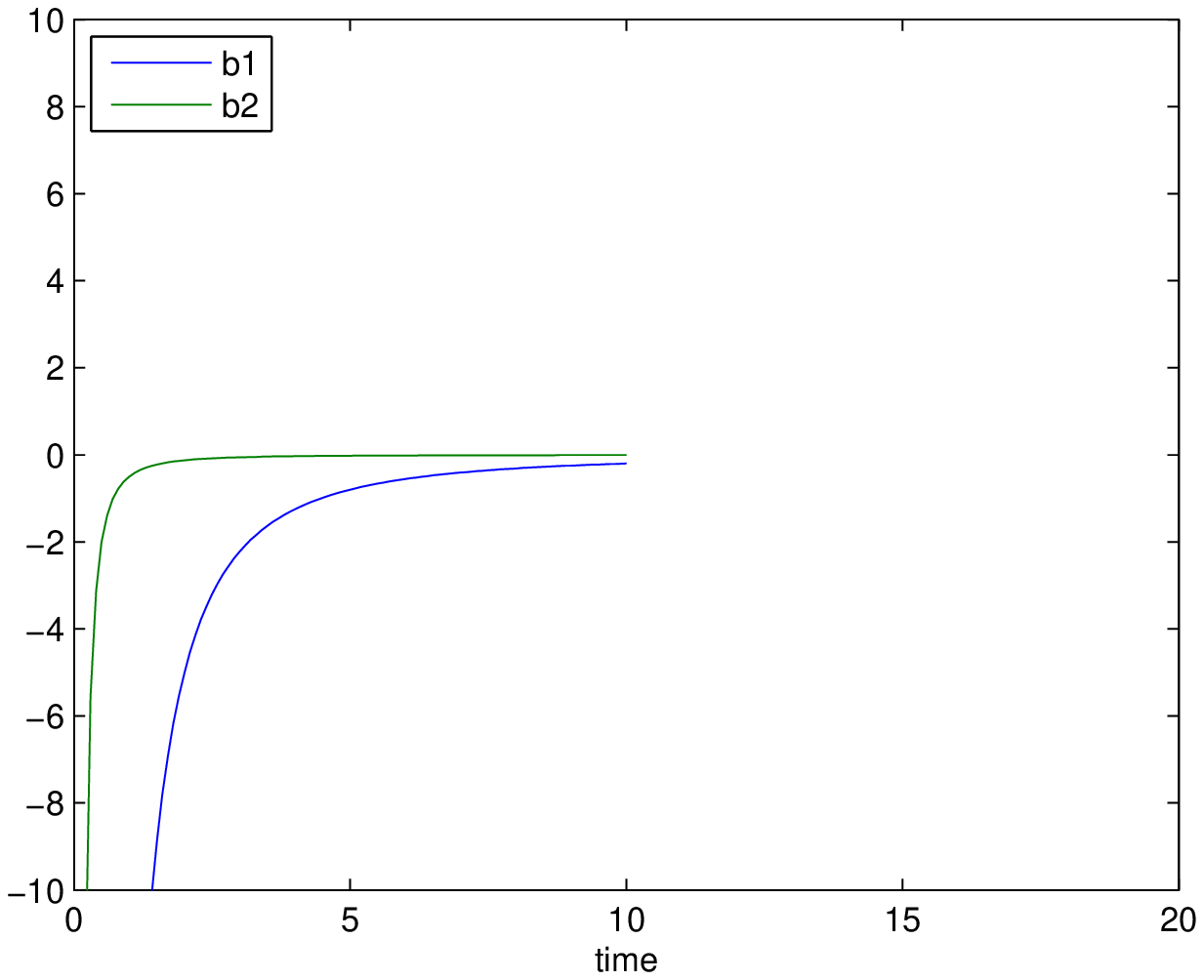}{0.45}{Variation of $\Pi$ for $\gamma=1$ with time
for the subcase Ideal Gas in Model 1. $b1$ and $b2$ represent
variation for the parameter $n=0.5$ and $n=2$ respectively.}
{0.45}{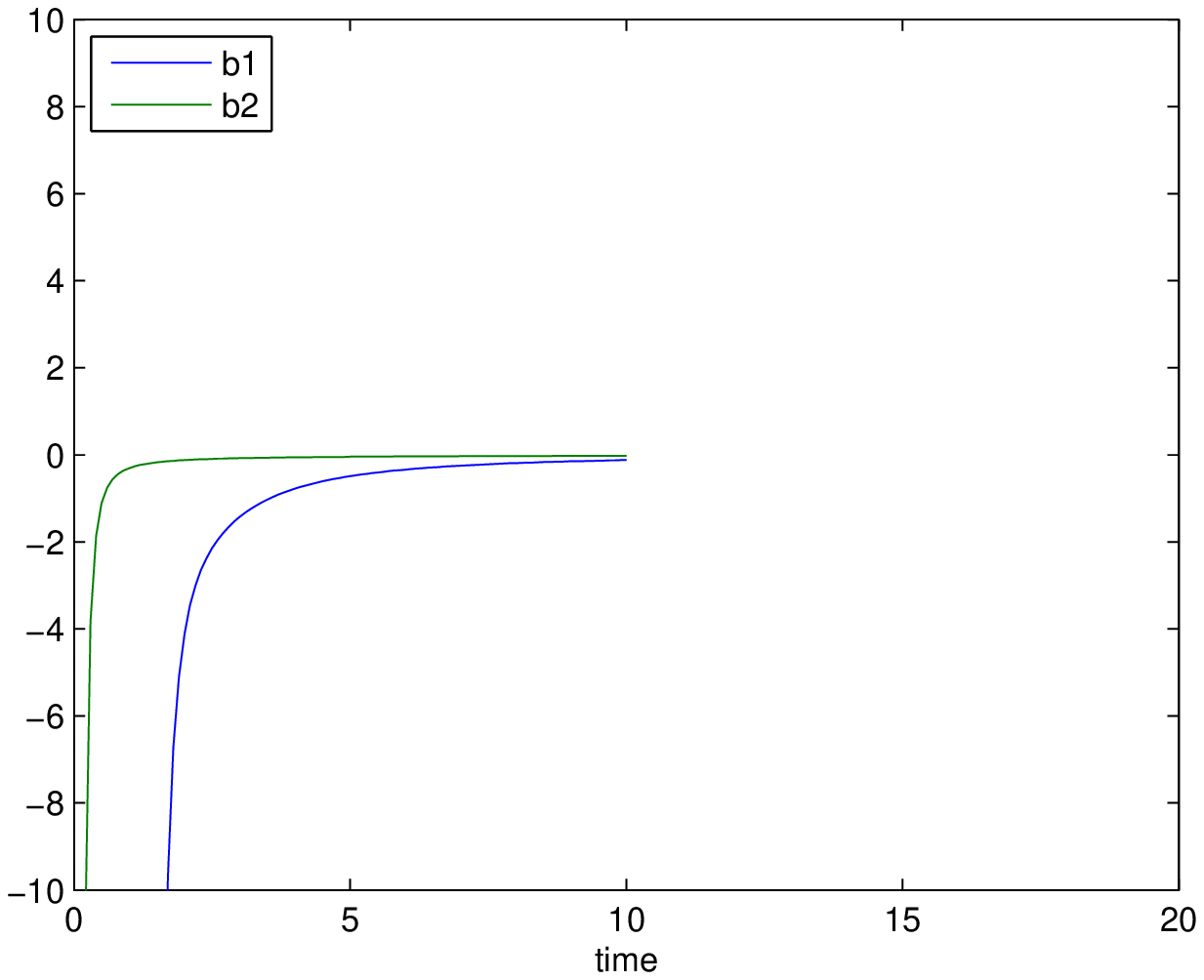}{0.45}{Variation of $\Pi$ for $\gamma=1/3$ with
time for the subcase Ideal Gas in Model 1.
 $b1$ and $b2$ represent variation for the parameter $n=0.5$ and $n=2$ respectively.}{0.45}

\noindent  In this case, the value of bulk viscous coefficient in
different theories can be obtained as

\noindent Truncated theory:\\
\begin{equation}
\zeta= \frac{\displaystyle
\begin{array}{c}\left[\zeta_{19}(nh_0t)^{-2}+\zeta_{20} (nh_0t)^{\frac{-6}{n}}-\zeta_{21}(nh_0t)^{-2/n}\right]\\
\left[\rho_0(nh_0t)^{-2}-\rho_1(nh_0t)^{\frac{-6}{n}}-\rho_2(nh_0t)^{\frac{-2}{n}}\right]\\
\end{array}}{\displaystyle
\begin{array}{c} \zeta_{22}(nh_0t)^{-3}+\zeta_{23}(nh_0t)^{\frac{-6}{n}-1}-\zeta_{24} (nh_0t)^{\frac{-2}{n}-1}\\
\end{array}}.
\end{equation}
Here $\zeta_{19}=\left[(1+\gamma)-\frac{2n}{3}\right]\rho_0$, $\zeta_{20}=(1-\gamma)\rho_1$,
$\zeta_{21}=\frac{(1+3\gamma)}{3}\rho_2$, $\zeta_{22}=\left[\left(1+\gamma-\frac{2n}{3}\right)2n+3\right]h_0 \rho_0$,
$\zeta_{23}=3(1-2\gamma)h_0 \rho_1$ and $\zeta_{24}=\frac{(11+6\gamma)}{3}h_0 \rho_2$.\\

\noindent  Full Causal theory:\\

\begin{equation}
\zeta= \frac{\displaystyle
\begin{array}{c}\left[\zeta_{25}(nh_0t)^{-2}+\zeta_{26} (nh_0t)^{\frac{-6}{n}}-\zeta_{27}(nh_0t)^{-2/n}\right]\\
\left[\rho_0(nh_0t)^{-2}-\rho_1(nh_0t)^{\frac{-6}{n}}-\rho_2(nh_0t)^{\frac{-2}{n}}\right]^{2}\\
\end{array}}{\displaystyle
\begin{array}{c} \left[\zeta_{28}(nh_0t)^{-3}+\zeta_{29}(nh_0t)^{\frac{-6}{n}-1}-\zeta_{30} (nh_0t)^{\frac{-2}{n}-1}\right]\\
\left[\rho_0(nh_0t)^{-2}-\rho_1(nh_0t)^{\frac{-6}{n}}-\rho_2(nh_0t)^{\frac{-2}{n}}\right]\\
+ 3h_0(nh_0t)^{-1}\left[\rho_0(nh_0t)^{-2}-\rho_1(nh_0t)^{\frac{-6}{n}}-\rho_2(nh_0t)^{\frac{-2}{n}}\right]^{2}\\
+\left[\zeta_{16}(nh_0t)^{-2}+\zeta_{17}(nh_0t)^{\frac{-6}{n}}+\zeta_{18}(nh_0t)^{\frac{-2}{n}}\right]\\
\left[\zeta_{31}(nh_0t)^{-3}-\zeta_{32}(nh_0t)^{\frac{-2}{n}-1}-\zeta_{33}(nh_0t)^{\frac{-6}{n}-1}\right]
\end{array}}
\end{equation}
where $\zeta_{25}=\left[(1+\gamma)-\frac{2n}{3}\right]\rho_0$, $\zeta_{26}=(1-\gamma)\rho_1$,
$\zeta_{27}=\frac{(1+3\gamma)}{3}\rho_2$, $\zeta_{28}=\left[(1+\gamma)-\frac{2n}{3}\right]2nh_0\rho_0$,
 $\zeta_{29}=6(1-\gamma)h_0\rho_1$, $\zeta_{30}=\frac{2(1+3\gamma)h_0\rho_2}{3}$,
  $\zeta_{31}=\left[\frac{3}{2}+\frac{n(1+2\gamma)}{(1+\gamma)}\right]\rho_0 h_0$,
  $\zeta_{32}=\left[\frac{3}{2}+\frac{(1+2\gamma)}{(1+\gamma)}\right]\rho_2 h_0$ and
   $\zeta_{33}=\left[\frac{3}{2}+\frac{3(1+2\gamma)}{(1+\gamma)}\right]\rho_1 h_0$.\\

\noindent  Eckart theory:\\
\begin{equation}
\zeta=\zeta_{34}(nh_0t)^{-1}+\zeta_{35}(nh_0t)^{\frac{-6}{n}+1}-\zeta_{36}(nh_0t)^{\frac{-2}{n}+1}.
\end{equation}
Here $\zeta_{34}=\left[(1+\gamma)-\frac{2n}{3}\right]\frac{\rho_0}{3h_0}$, $\zeta_{35}=\frac{(1-\gamma)\rho_1}{3h_0}$
and $\zeta_{36}=\frac{(1+3\gamma)\rho_2}{9h_0}$.
\subsubsection{Creation with Second Order Correction in $H$}
The conservation of total particle number in standard cosmology can be generalized by using the Taylor expansion of
 $\frac{\dot{\eta}}{\eta}=f(H)$ up to second order in $H$. This idea was given by
 Triginer and Pavon in 1994 \cite{Triginer}. So here we have
\begin{equation}
\frac{\dot{\eta}}{\eta}=-3H+dH^2
\end{equation}
where $d$ is a constant. The particle number density $\eta$ also satisfies the balance equation
\begin{equation}
\dot{\eta}+3\eta H=\Gamma.
\end{equation}
So from above equations
\begin{equation}
\Gamma=d\eta H^2.
\end{equation}
Here creation, no creation and annihilation of particles are decided by the conditions $d>0$, $d=0$ and $d<0$ respectively.
Using equation (69), creation pressure takes the form
\begin{equation}
p_c=-\frac{(1+\gamma)d}{3}H\rho.
\end{equation}
Hence the expression for creation pressure and bulk viscous stress can be given as
\begin{equation}
p_c=-\frac{(1+\gamma)d h_0}{3}\left[\rho_0(nh_0t)^{-3}-\rho_1 (nh_0t)^{\frac{-6}{n}-1}-\rho_2(nh_0t)^{\frac{-2}{n}-1}\right]
\end{equation}
\begin{eqnarray}
\Pi &=& \zeta_{37}(nh_0t)^{-2}-\zeta_{38}(nh_0t)^{\frac{-6}{n}}+\zeta_{39}(nh_0t)^{\frac{-2}{n}}\nonumber \\
&&
+\zeta_{40}(nh_0t)^{-3}-\zeta_{41}(nh_0t)^{\frac{-6}{n}-1}-\zeta_{42}(nh_0t)^{\frac{-2}{n}-1}.
\end{eqnarray}
Here $\zeta_{37}=\left[\frac{2n}{3}-(1+\gamma)\right]\rho_0$, $\zeta_{38}=(1-\gamma)\rho_1$, $\zeta_{39}=\gamma$,
 $\zeta_{40}=\frac{d(1+\gamma)h_0 \rho_0}{3}$, $\zeta_{41}=\frac{d(1+\gamma)h_0 \rho_1}{3}$ and
  $\zeta_{42}=\frac{d(1+\gamma)h_0 \rho_2}{3}$. The behaviour of $Pc$ and $\Pi$ in this case can be observed
 in the following figures from 17 to 22.\\

\myfigures{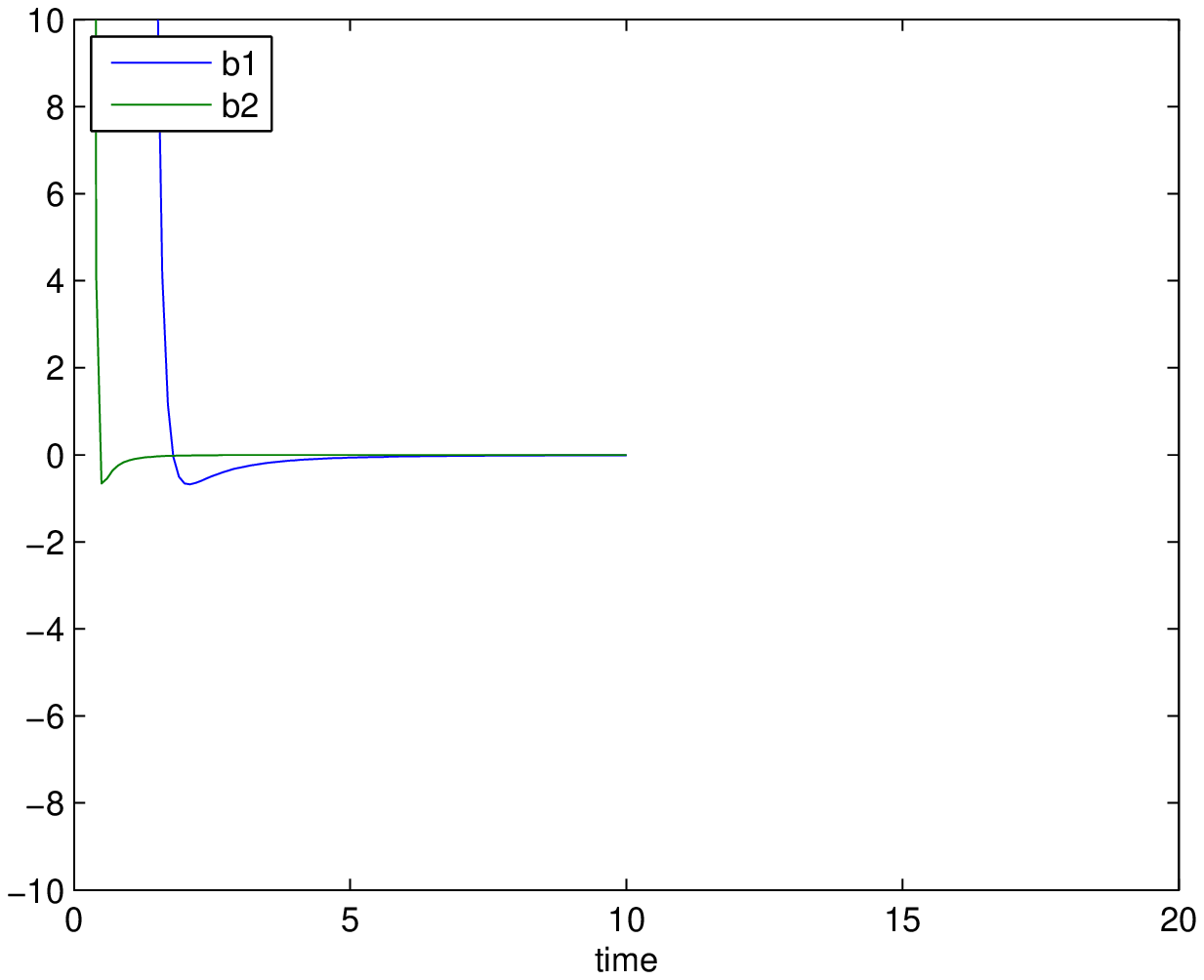}{0.45}{Variation of creation pressure $Pc$ for
$\gamma=0$ with time for the subcase Creation with Second Order
Correction in $H$ in Model 1. $b1$ and $b2$ represent variation for
the parameter $n=0.5$ and $n=2$ respectively.}
{0.45}{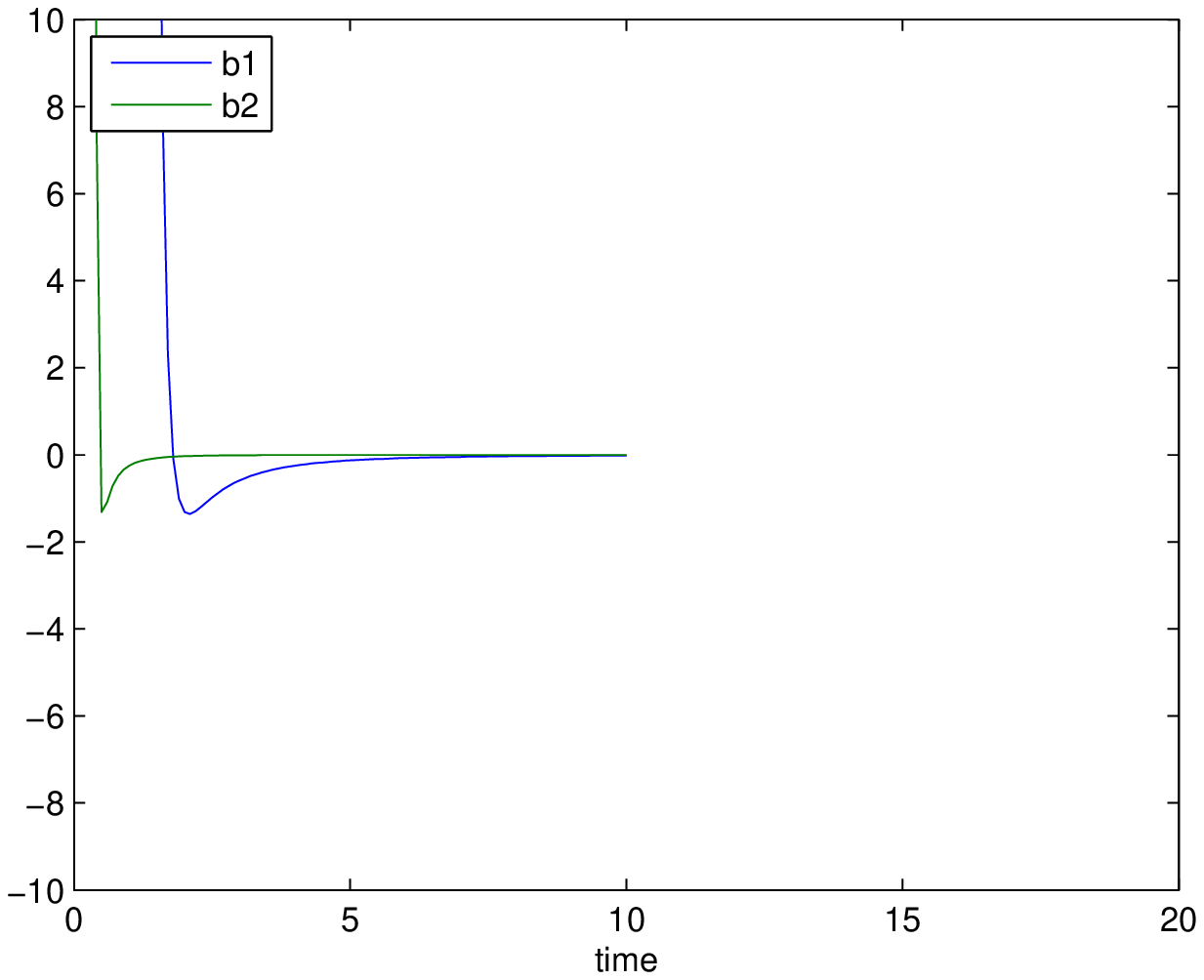}{0.45}{Variation of creation pressure $Pc$ for
$\gamma=1$ with time for the subcase Creation with Second Order
 Correction in $H$ in Model 1. $b1$ and $b2$ represent variation for the parameter $n=0.5$ and $n=2$ respectively.}{0.45}

\myfigures{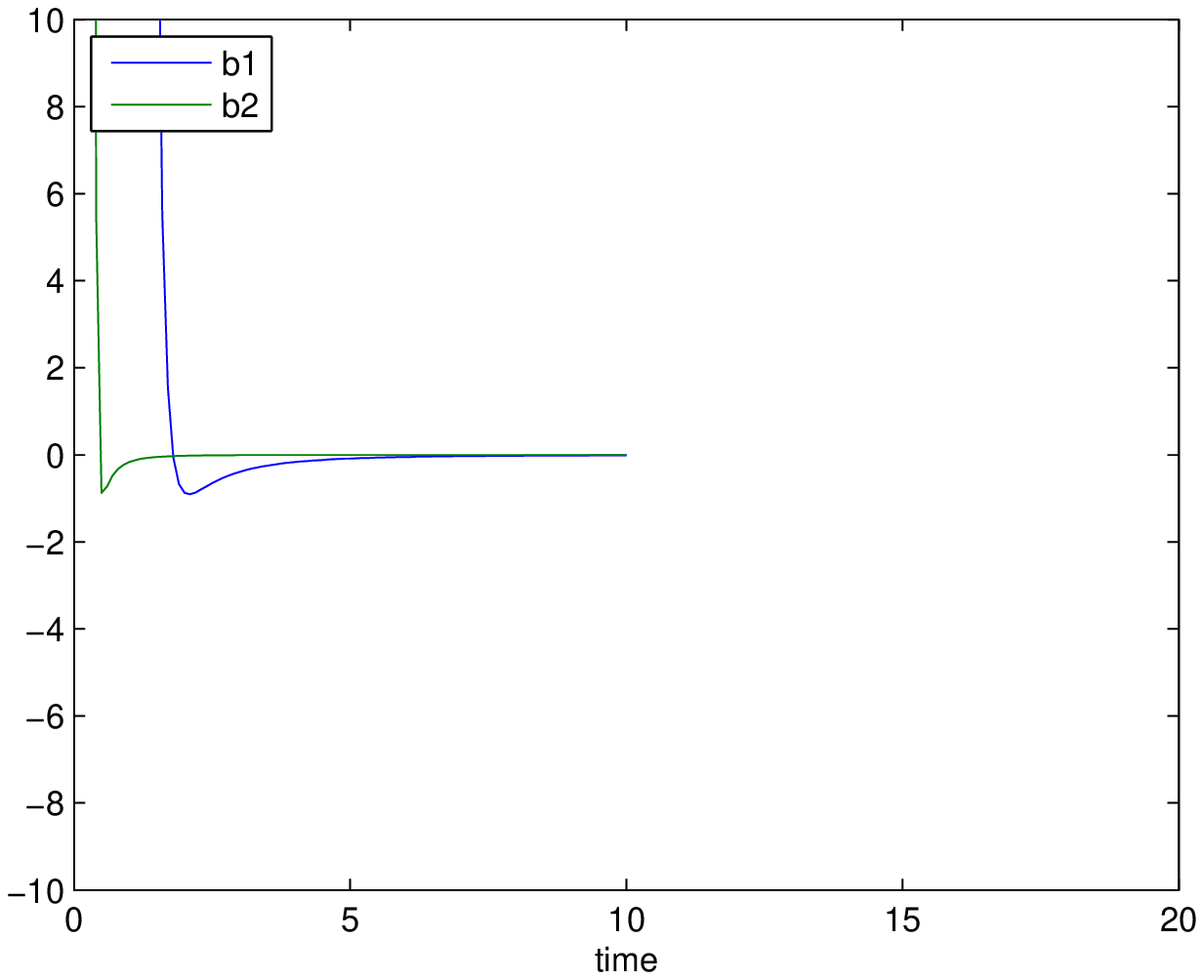}{0.45}{Variation of creation pressure $Pc$ for
$\gamma=1/3$ with time for the subcase Creation with Second Order
 Correction in $H$ in Model 1. $b1$ and $b2$ represent variation for the parameter $n=0.5$ and $n=2$ respectively.}
{0.45}{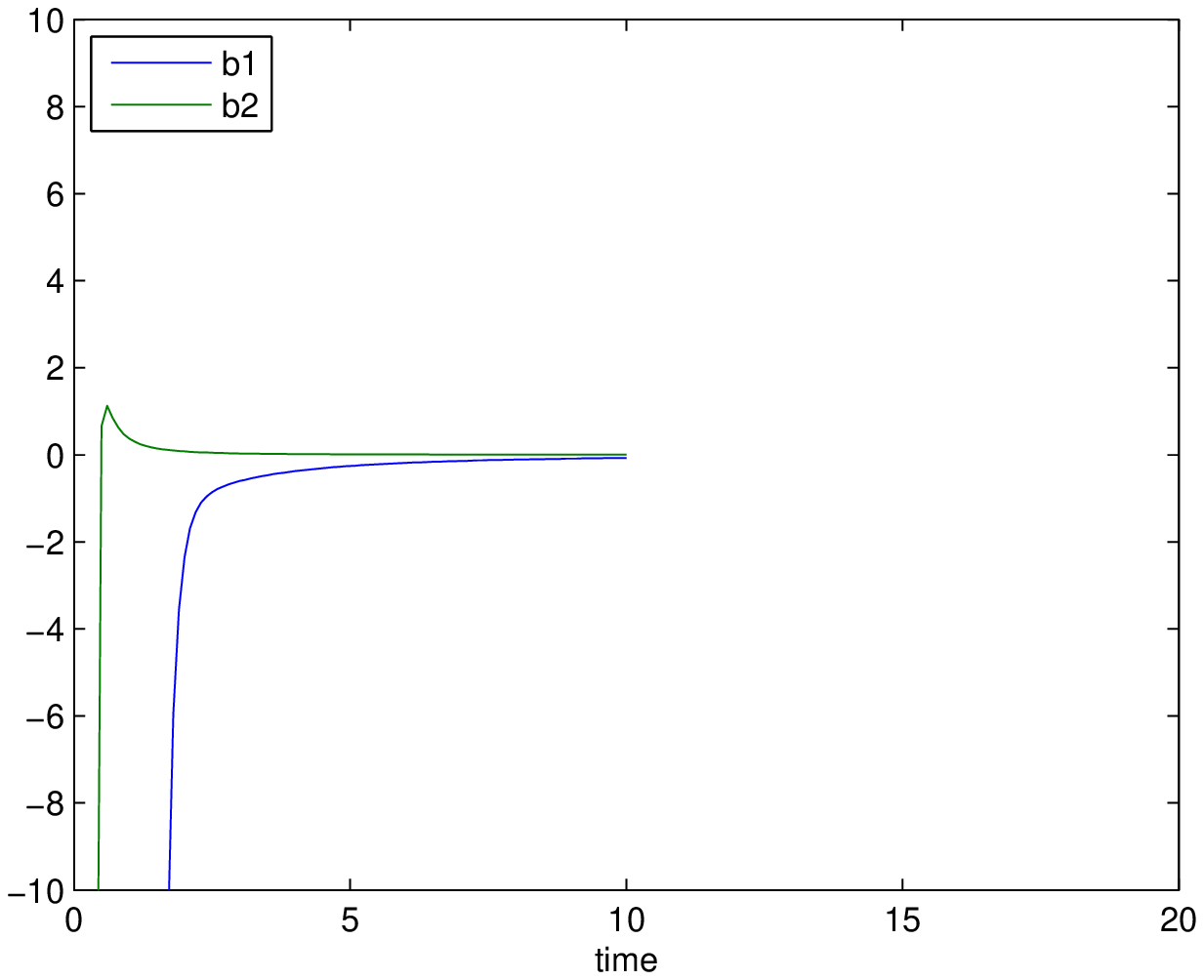}{0.45}{Variation of $\Pi$ for $\gamma=0$ with time for
the subcase Creation with Second Order Correction in $H$ in Model 1.
$b1$ and $b2$ represent variation for the parameter $n=0.5$ and
$n=2$ respectively.}{0.45}

\myfigures{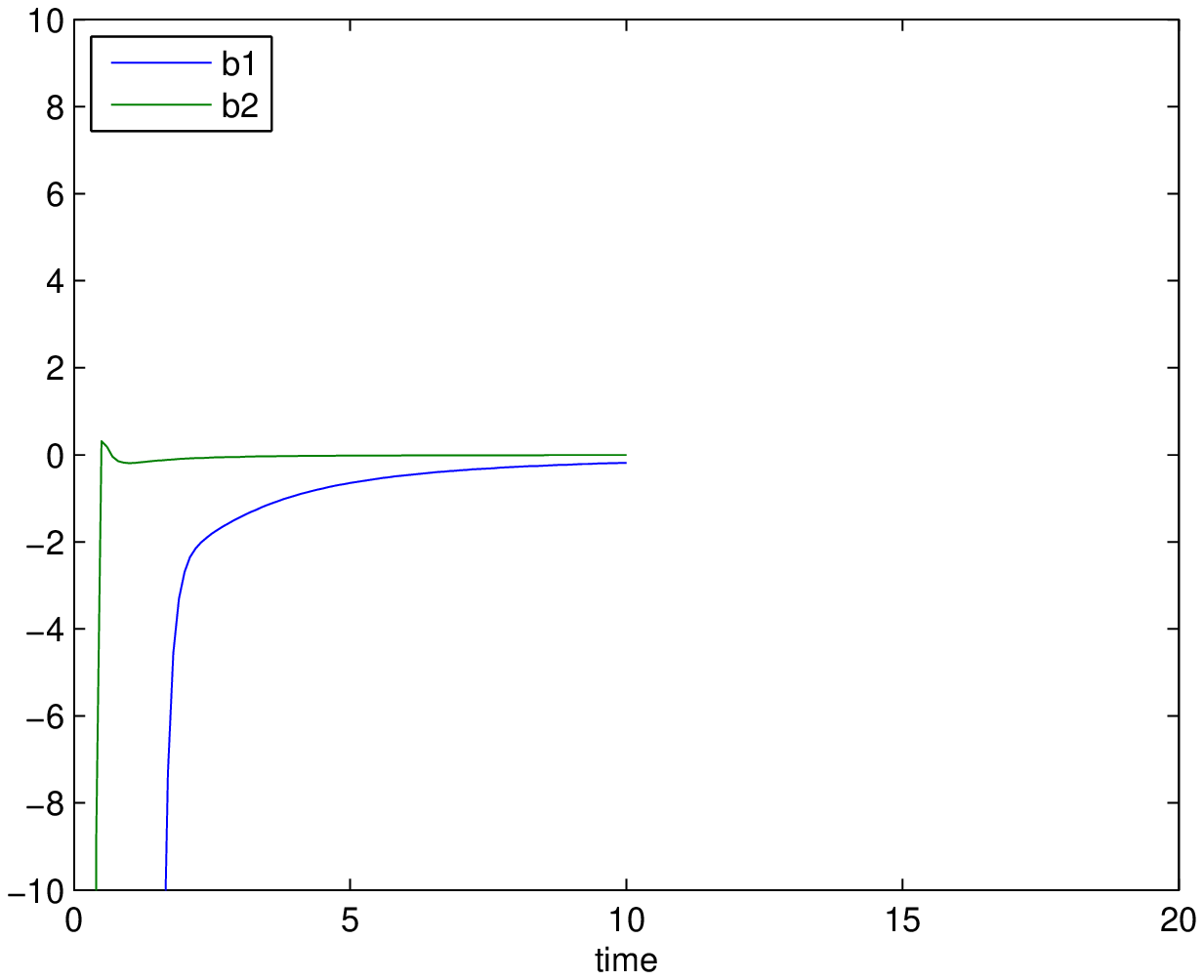}{0.45}{Variation of $\Pi$ for $\gamma=1$ with time
for the subcase Creation with Second Order Correction in $H$ in
Model 1. $b1$ and $b2$ represent variation for the parameter $n=0.5$
and $n=2$ respectively.} {0.45}{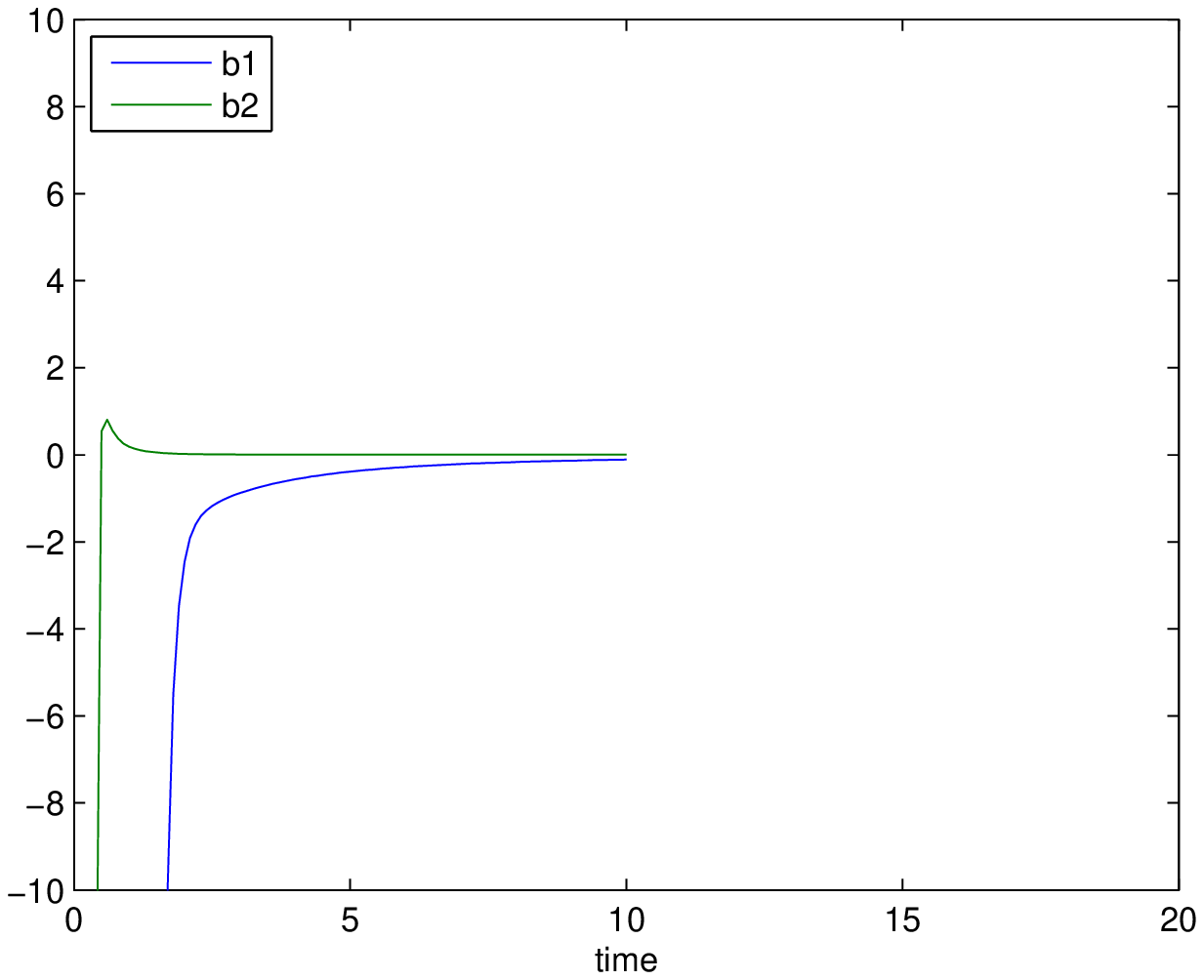}{0.45}{Variation of $\Pi$
for $\gamma=1/3$ with time for the subcase Creation with Second
Order Correction in $H$ in Model 1. $b1$ and $b2$ represent
variation for the parameter $n=0.5$ and $n=2$ respectively.}{0.45}

\noindent In this case, the value of bulk viscous coefficient in different theories can be obtained as\\

\noindent  Truncated theory:\\

\begin{equation}
\zeta= \frac{\displaystyle
\begin{array}{c}[-\zeta_{37}(nh_0t)^{-2}+\zeta_{38}(nh_0t)^{\frac{-6}{n}}-\zeta_{39}(nh_0t)^{\frac{-2}{n}}\\
-\zeta_{40}(nh_0t)^{-3}+\zeta_{41}(nh_0t)^{\frac{-6}{n}-1}+\zeta_{42}(nh_0t)^{\frac{-2}{n}-1}]\\
\times[\rho_0(nh_0t)^{-2}-\rho_1(nh_0t)^{\frac{-6}{n}}-\rho_2(nh_0t)^{\frac{-2}{n}}]
\end{array}}{\displaystyle
\begin{array}{c} [\zeta_{43}(nh_0t)^{-3}+\zeta_{44}(nh_0t)^{\frac{-6}{n}-1}-\zeta_{45}(nh_0t)^{\frac{-2}{n}-1}\\
-\zeta_{46}(nh_0t)^{-4}+\zeta_{47}(nh_0t)^{\frac{-6}{n}-2}+\zeta_{48}(nh_0t)^{\frac{-2}{n}-2}]
\end{array}}
\end{equation}
where $\zeta_{43}=\left[2n(1+\gamma)+3-\frac{4n^2}{3}\right]$, $\zeta_{44}=3(1-2\gamma)\rho_1 h_0$,
 $\zeta_{45}=(2\gamma+3)h_0 \rho_2$, $\zeta_{46}=d(1+\gamma)h_0^{2}\rho_0 n$,
  $\zeta_{47}=\left[\frac{(n+6)d(1+\gamma)h_0^{2}\rho_1}{3}\right]$ and
  $\zeta_{48}=\left[\frac{d(n+2)(1+\gamma)h_0^{2}\rho_2}{3}\right]$.\\

\noindent  Full Causal theory:\\

\begin{equation}
\zeta= \frac{\displaystyle
\begin{array}{c}[-\zeta_{37}(nh_0t)^{-2}+\zeta_{38}(nh_0t)^{\frac{-6}{n}}-\zeta_{39}(nh_0t)^{\frac{-2}{n}}\\
-\zeta_{40}(nh_0t)^{-3}+\zeta_{41}(nh_0t)^{\frac{-6}{n}-1}+\zeta_{42}(nh_0t)^{\frac{-2}{n}-1}]\\
\times [\rho_0(nh_0t)^{-2}-\rho_1(nh_0t)^{\frac{-6}{n}}-\rho_2(nh_0t)^{\frac{-2}{n}}]^{2}
\end{array}}{\displaystyle
\begin{array}{c} [\zeta_{49}(nh_0t)^{-3}+\zeta_{50}(nh_0t)^{\frac{-6}{n}-1}-\zeta_{51}(nh_0t)^{\frac{-2}{n}-1}\\
-\zeta_{52}(nh_0t)^{-4}+\zeta_{53}(nh_0t)^{\frac{-6}{n}-2}+\zeta_{54}(nh_0t)^{\frac{-2}{n}-2}]\\
\times [\rho_0(nh_0t)^{-2}-\rho_1(nh_0t)^{\frac{-6}{n}}-\rho_2(nh_0t)^{\frac{-2}{n}}]\\
+3h_0(nh_0t)^{-1}[\rho_0(nh_0t)^{-2}-\rho_1(nh_0t)^{\frac{-6}{n}}-\rho_2(nh_0t)^{\frac{-2}{n}}]^{2}\\
+[\zeta_{37}(nh_0t)^{-2}-\zeta_{38}(nh_0t)^{\frac{-6}{n}}+\zeta_{39}(nh_0t)^{\frac{-2}{n}}\\
+\zeta_{40}(nh_0t)^{-3}-\zeta_{41}(nh_0t)^{\frac{-6}{n}-1}-\zeta_{42}(nh_0t)^{\frac{-2}{n}-1}]\\
\times [\zeta_{55}(nh_0t)^{-3}-\zeta_{56}(nh_0t)^{\frac{-6}{n}-1}-\zeta_{57}(nh_0t)^{\frac{-2}{n}-1}]
\end{array}}
\end{equation}
where $\zeta_{49}=\left[2(1+\gamma)-\frac{4n}{3}\right]n h_0\rho_0$, $\zeta_{50}=6(1-\gamma)\rho_1 h_0$,
 $\zeta_{51}=2h_0 \rho_2 \gamma$, $\zeta_{52}=d(1+\gamma)h_0^{2}\rho_0 n$,
  $\zeta_{53}=\left[\frac{(n+6)d(1+\gamma)h_0^{2}\rho_1}{3}\right]$,
  $\zeta_{54}=\left[\frac{d(n+2)(1+\gamma)h_0^{2}\rho_2}{3}\right]$,
  $\zeta_{55}=\left[\frac{3}{2} h_0-\frac{(1+2\gamma)}{2(1+\gamma)}\right]\rho_0$,
   $\zeta_{56}=\left[\frac{3}{2} +\frac{3(1+2\gamma)}{(1+\gamma)}\right]\rho_1 h_0$,
    $\zeta_{57}=\left[\frac{3}{2} +\frac{(1+2\gamma)}{(1+\gamma)}\right] h_0 \rho_2$. \\

\noindent  Eckart theory:\\
\begin{equation}
\zeta= \frac{\displaystyle
\begin{array}{c}[-\zeta_{37}(nh_0t)^{-2}+\zeta_{38}(nh_0t)^{\frac{-6}{n}}-\zeta_{39}(nh_0t)^{\frac{-2}{n}}\\
-\zeta_{40}(nh_0t)^{-3}+\zeta_{41}(nh_0t)^{\frac{-6}{n}-1}+\zeta_{42}(nh_0t)^{\frac{-2}{n}-1}]
\end{array}}{\displaystyle
\begin{array}{c} 3h_0(nh_0t)^{-1}
\end{array}}.
\end{equation}
\subsection{Model 2}
In this case, putting the value of the average scale-factor from equation (33) into equations (34)-(36), the exact values of
 the metric functions can be obtained as
\begin{equation}
A=a_0 \exp{(h_0t)}
\end{equation}
\begin{equation}
B=B_0 a_0 \exp{\left[h_0t-\frac{X}{3a_1h_0}\exp{(-h_0t)}\right]}
\end{equation}
\begin{equation}
C=C_0 a_0 \exp{\left[h_0t+\frac{X}{3a_1h_0}\exp{(-h_0t)}\right]}.
\end{equation}

Hence the Expansion scalar $\theta$, Shear scalar $\sigma^2$, generalized Hubble parameter $H$ and the Volume scalar $V$ can be
 written as
\begin{equation}
\theta=3h_0
\end{equation}
\begin{equation}
\sigma^2=\frac{X^2}{a_1^2}\exp{(-2h_0t)}
\end{equation}
\begin{equation}
H=h_0
\end{equation}
and
\begin{equation}
V=V_0\exp{(3h_0t)}.
\end{equation}
The anisotropy parameter $A_m$ can be given by
\begin{equation}
A_m=\frac{2}{3}\frac{X^2}{h_0^{2}a_1^{2}}\exp{(-2h_0t)}.
\end{equation}
The directional Hubble parameters can be obtained as
\begin{equation}
H_1=h_0
\end{equation}
\begin{equation}
H_2=h_0+\frac{X}{3a_1}\exp{(-h_0t)}
\end{equation}
\begin{equation}
H_3=h_0-\frac{X}{3a_1}\exp{(-h_0t)}.
\end{equation}
Here the value of the energy density can be calculated as
\begin{equation}
\rho=\rho_3-\rho_6 \exp{(-2h_0t)}
\end{equation}
where $\rho_3=\frac{3h_0^{2}}{8\pi G}$, $\rho_6=\rho_4+\rho_5$,
$\rho_4=\frac{X^2}{72\pi G a_1^{2}}$, $\rho_5=\frac{3m^2}{8\pi G
a_1^{2/3}}$. So, we have the equation for pressure as
\begin{equation}
p=\gamma\left[\rho_3-\rho_6 \exp{(-2h_0t)}\right].
\end{equation}
Here we observe that all three scale factors $A$, $B$, $C$, the
shear scalar $\sigma^{2}$, the volume scalar $V$, the anisotropy
parameter $A_m$, the energy density $\rho$ and the pressure $p$ are
all constant at the initial time $t=0$. The variation of $A$, $B$
and $C$ can be observed graphically in figure 23. The expansion
scalar $\theta$ and the generalized Hubble parameter $H$ are
constants through out the evolution of the universe. The directional
Hubble parameters $H_1$, $H_2$ and $H_3$ are constant at this time
$t=0$. These results show that the universe starts evolving with
constant volume and expands with exponential rate. Thus the model
represents uniform expansion due to the constant expansion scalar
and volume grows exponentially with time. Similarly we can study
this model for very large time as $t \rightarrow \infty $. We find
that for large time all three scale factors and the volume scalar
will become infinity. The mean anisotropy parameter and shear scalar
decrease with time and tend to zero at $t \rightarrow \infty $ which
means that the universe is accelerating in later stage of its
evolution. Also the directional Hubble parameters become constant
and uniform. Also the energy density so as pressure become
constant for large time.\\

\noindent Now in this case also we study in the following subsections the behaviour of particle creation and bulk
 viscosity of this model in four different physical laws. That will give four different sub models out of the above model.
\subsubsection{Bulk Viscosity Energy-Density Law}
The bulk viscous stress and creation pressure in this case can be obtained as
\begin{equation}
\Pi=\Pi_0[\rho_3-\rho_6 \exp{(-2h_0t)}]^{\omega}.
\end{equation}
\begin{equation}
p_c=p_3+p_4 \exp{(-2h_0t)}-\Pi_0 [\rho_3-\rho_6 \exp{(-2h_0t)}]^{\omega}.
\end{equation}
Here $p_3=-(1+\gamma)\rho_3$, $p_4=\left[(1+\gamma)\rho_6-\frac{2}{3}\rho_6\right]$. The following figures from
24 to 27 show the variation of $\Pi$ and $Pc$ in this case for Model 2. \\

\myfigures{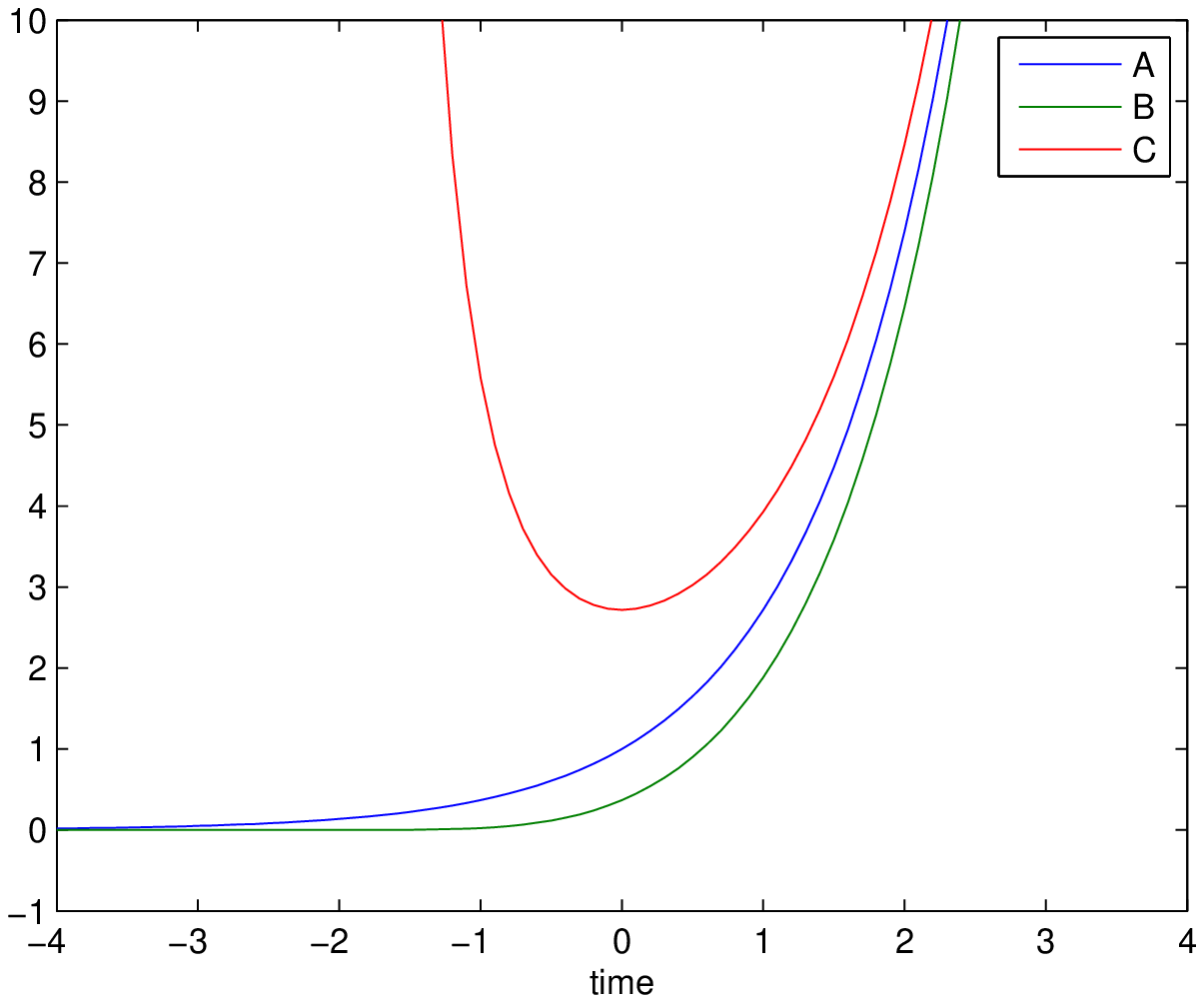}{0.45}{Variation of scale factors $A$, $B$ and $C$
with time for the Model 2.} {0.45}{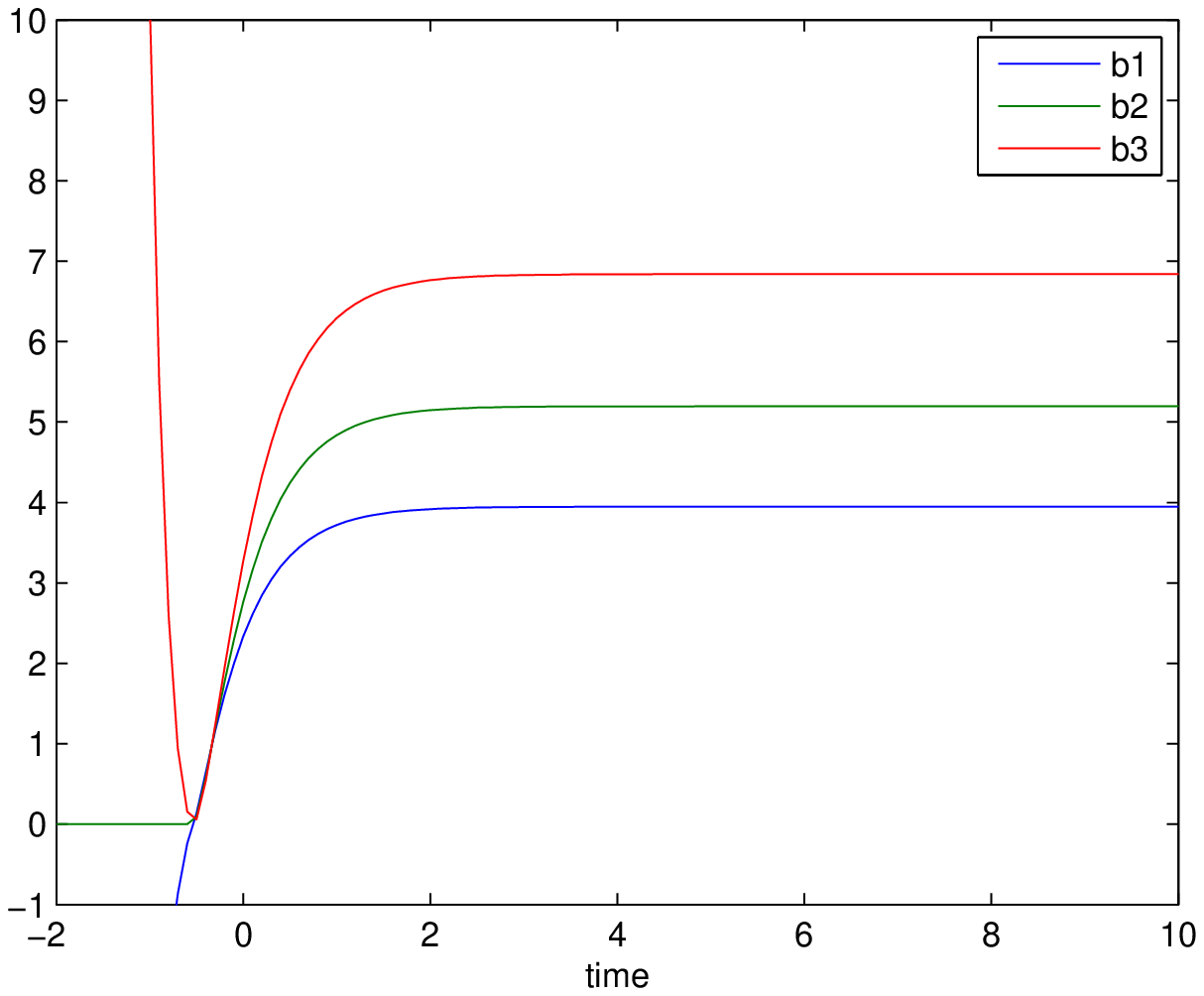}{0.45}{Variation of $\Pi$
with time for the subcase Bulk Viscosity Energy-Density Law in Model
2. $b1$, $b2$ and $b3$ represent the variation for $\omega=1.25$,
$\omega=1.5$ and $\omega=1.75$ respectively.}{0.45}

\myfigures{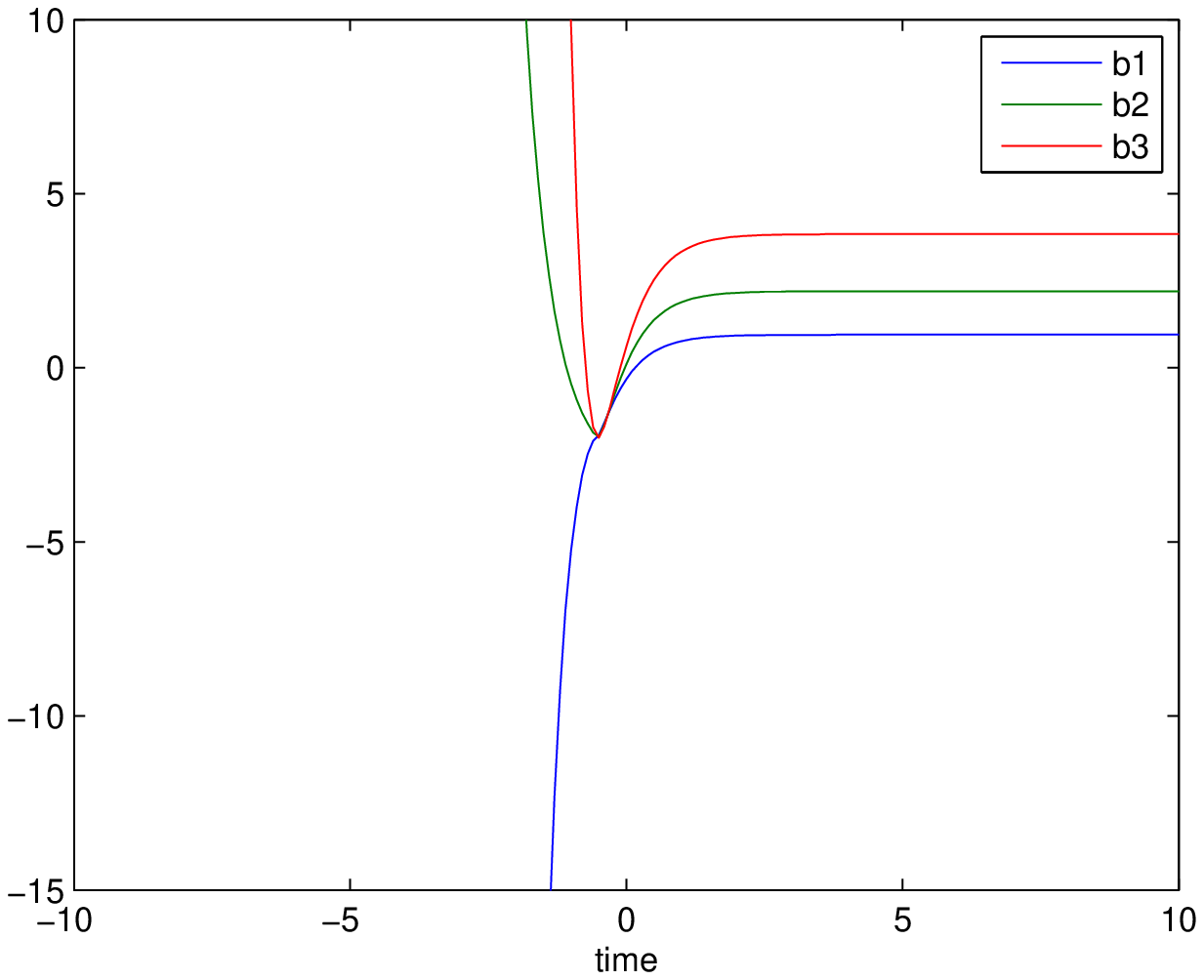}{0.45}{Variation of creation pressure $Pc$ for
$\gamma=0$ with time for the subcase Bulk Viscosity Energy-Density
Law in Model 2. $b1$, $b2$ and $b3$ represent the variation for
$\omega=1.25$, $\omega=1.5$ and $\omega=1.75$ respectively.}
{0.45}{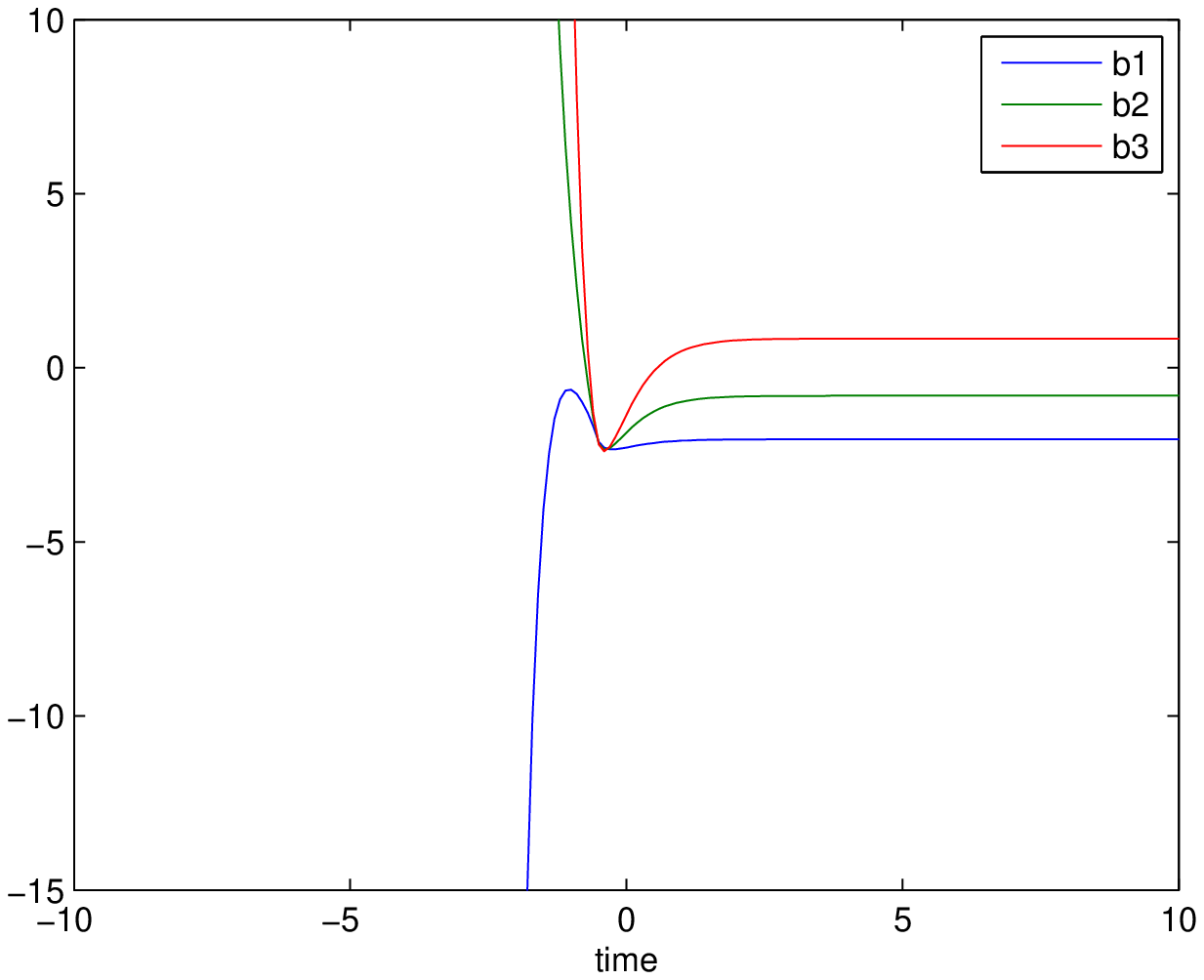}{0.45}{Variation of creation pressure $Pc$ for
$\gamma=1$ with time for the subcase Bulk Viscosity
 Energy-Density Law in Model 2. $b1$, $b2$ and $b3$ represent the variation for $\omega=1.25$,
 $\omega=1.5$ and $\omega=1.75$ respectively.}{0.45}

\noindent The bulk viscosity coefficient in three different theories can be given as\\

\noindent Truncated theory:\\
\begin{equation}
\zeta= \frac{\displaystyle
\begin{array}{c} -\Pi_0 [\rho_3-\rho_6 \exp{(-2h_0t)}]^{\omega+1}
\end{array}}{\displaystyle
\begin{array}{c} \zeta_{58}-\zeta_{59}\exp{(-2h_0t)}+\zeta_{60}\exp{(-2h_0t)}[\rho_3-\rho_6 \exp{(-2h_0t)}]^{\omega-1}
\end{array}}
\end{equation}
where $\zeta_{58}=3 h_0\rho_3$, $\zeta_{59}=3h_0\rho_{6}$ and $\zeta_{60}=2h_0\Pi_0 \omega \rho_{6}$.\\

\noindent Full Causal theory:\\
\begin{equation}
\zeta= \frac{\displaystyle
\begin{array}{c} -2\Pi_0 [\rho_3-\rho_6 \exp{(-2h_0t)}]^{\omega+2}
\end{array}}{\displaystyle
\begin{array}{c} \zeta_{61}\exp{(-2h_0t)}[\rho_3-\rho_6\exp{(-2h_0t)}]^{\omega}+\zeta_{62}[\rho_3-\rho_6 \exp{(-2h_0t)}]^{\omega+1}\\
+6h_0[\rho_3-\rho_6 \exp{(-2h_0t)}]^{2}-\zeta_{63}\exp{(-2h_0t)}
\end{array}}
\end{equation}
where $\zeta_{61}=4 h_0\rho_6 \omega \Pi_0$, $\zeta_{62}=3\Pi_0 h_0$ and $\zeta_{63}=\frac{2(1+2\gamma)h_0\rho_6}{(1+\gamma)}$.\\

\noindent Eckart theory:\\
\begin{equation}
\zeta=-\frac{\Pi_0}{3h_0}[\rho_3-\rho_6 \exp{(-2h_0t)}]^{\omega}.
\end{equation}
\subsubsection{Uniform Particle Number Density $(\dot{\eta}=0)$}
The bulk viscous stress and creation pressure can be obtained as follows:
\begin{equation}
\Pi=-\frac{2}{3}\rho_6\exp{(-2h_0t)}
\end{equation}
and
\begin{equation}
p_c=-(1+\gamma)[\rho_3-\rho_6 \exp{(-2h_0t)}].
\end{equation}
The behavior of $\Pi$ and $Pc$ for Uniform Particle Number Density
in Model2 are presented graphically in figures 28 and 29.\\

\myfigures{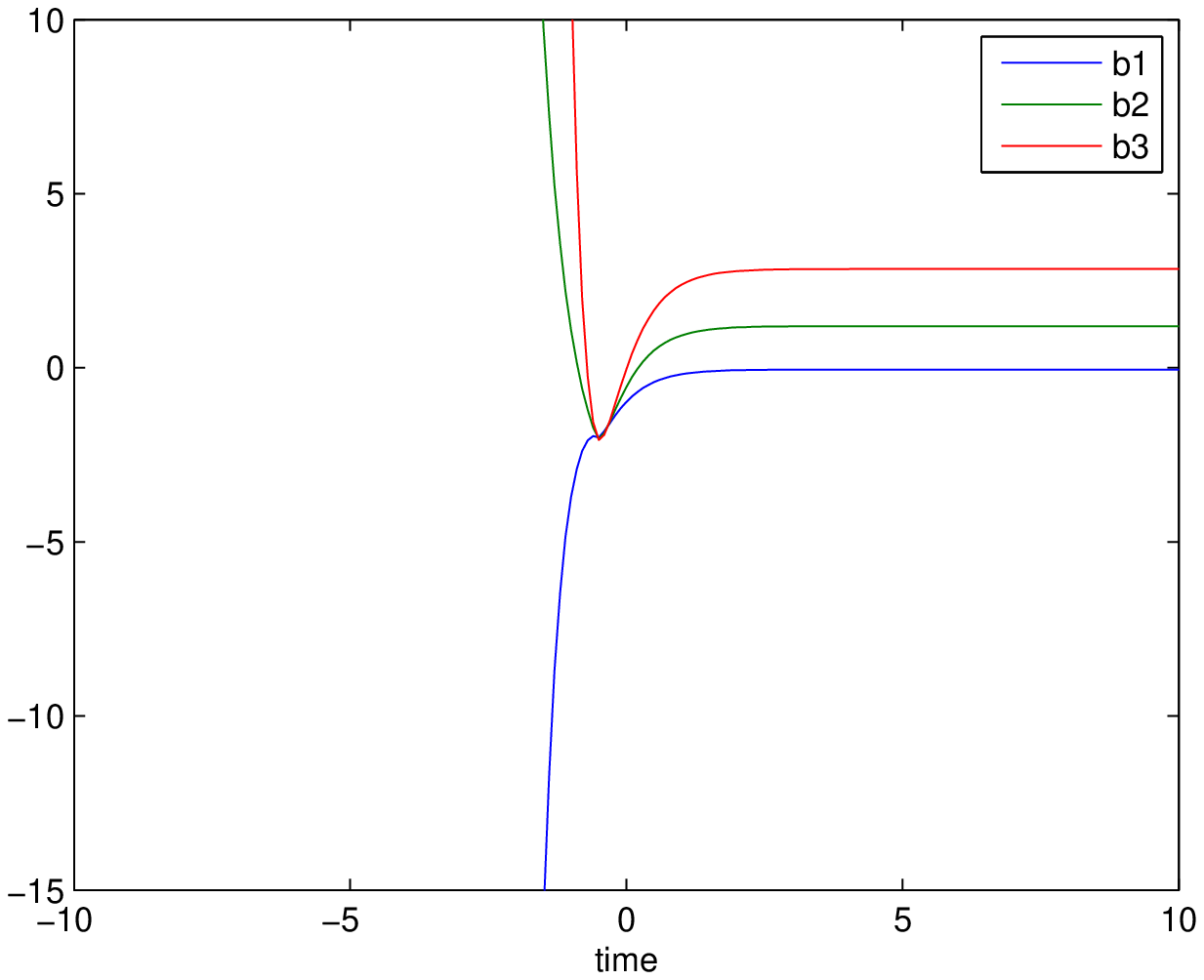}{0.45}{Variation of creation pressure $Pc$ for
$\gamma=1/3$ with time for the subcase Bulk Viscosity Energy-Density
Law in Model 2. $b1$, $b2$ and $b3$ represent the variation for
$\omega=1.25$, $\omega=1.5$ and $\omega=1.75$ respectively.}
{0.45}{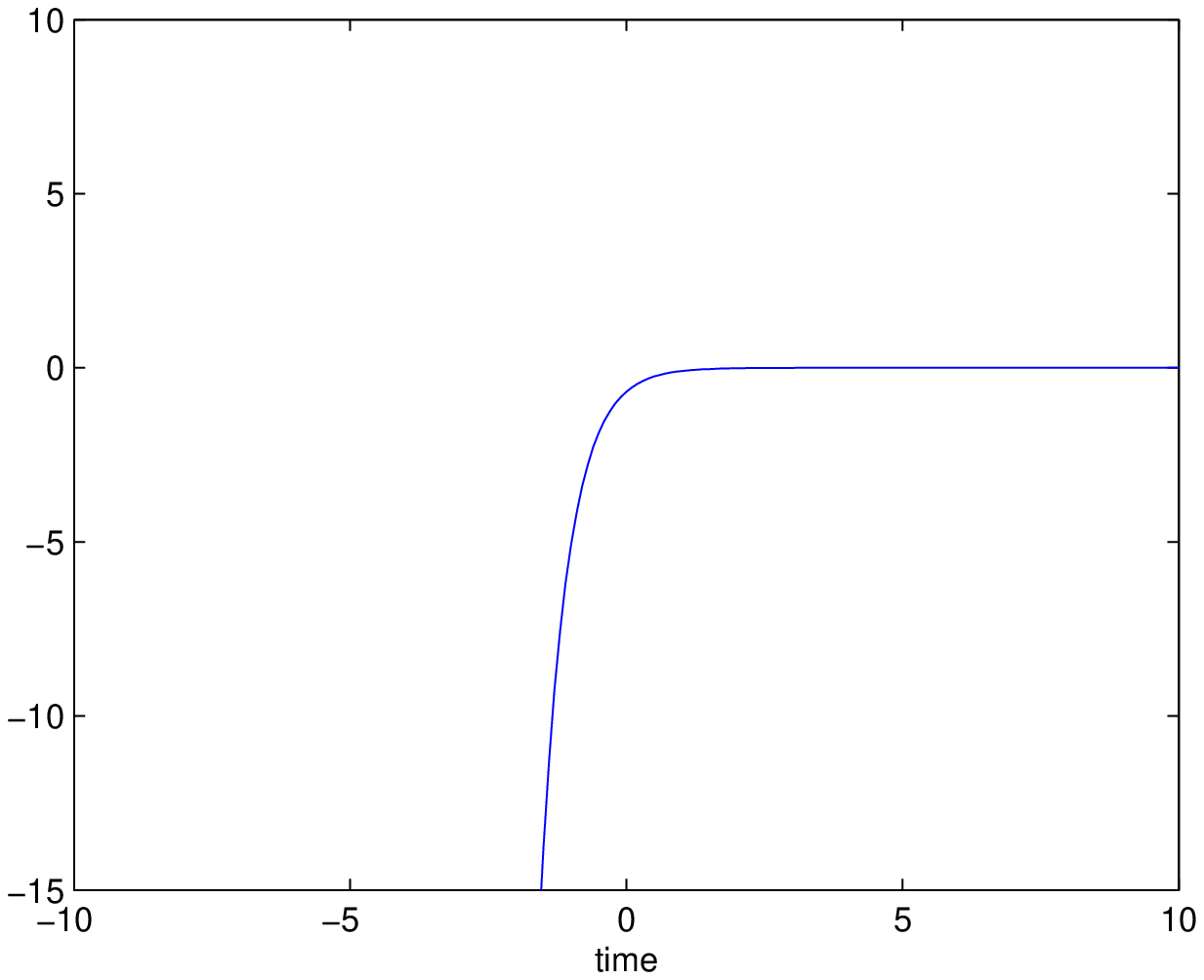}{0.45}{Variation of $\Pi$ with time for the subcase
Uniform Particle Number Density in Model 2.}{0.45}

\noindent The bulk viscosity coefficient in different theories can be written as\\

\noindent Truncated theory:\\
\begin{equation}
\zeta= \frac{\displaystyle
\begin{array}{c} \zeta_{64} \exp{(-2h_0t)}[\rho_3-\rho_6 \exp{(-2h_0t)}]
\end{array}}{\displaystyle
\begin{array}{c} \zeta_{65}-\zeta_{66}\exp{(-2h_0t)}
\end{array}}
\end{equation}
where $\zeta_{64}=\frac{2}{3}\rho_6$, $\zeta_{65}=3h_0 \rho_3$ and $\zeta_{66}=\frac{5}{3}h_0\rho_6$.\\

\noindent Full Causal theory:\\
\begin{equation}
\zeta= \frac{\displaystyle
\begin{array}{c} \zeta_{67} \exp{(-2h_0t)}[\rho_3-\rho_6 \exp{(-2h_0t)}]^{2}
\end{array}}{\displaystyle
\begin{array}{c} \zeta_{68}\exp{(-2h_0t)}+\zeta_{69}\exp{(-4h_0t)}
\end{array}}
\end{equation}
where $\zeta_{67}=\frac{2}{3}\rho_6$, $\zeta_{68}=-\frac{17}{3}h_0\rho_3 \rho_6$ and
 $\zeta_{69}=\left[\frac{2(1+2\gamma)}{3(1+\gamma)}+\frac{8}{3}\right]h_0\rho_6^{2}$.\\

\noindent Eckart theory:\\
\begin{equation}
\zeta=\zeta_{70}\exp{(-2h_0t)}.
\end{equation}
Here $\zeta_{70}=\frac{2\rho_6}{9h_0}$.
\subsubsection{Ideal Gas}
The value of the particle number density and Bulk viscous stress in this case can be obtained as
\begin{equation}
\eta=\eta_0 \exp{(-3h_0t)}
\end{equation}
and
\begin{equation}
\Pi=-\Pi_1+\Pi_2\exp{(-2h_0t)}
\end{equation}
where $\Pi_1=(1+\gamma)\rho_3$, $\Pi_2=\frac{(1+3\gamma)}{3}\rho_6$. The behaviour of $\Pi$ in Ideal Gas in this model
can be observed in figure 30. \\

\myfigures{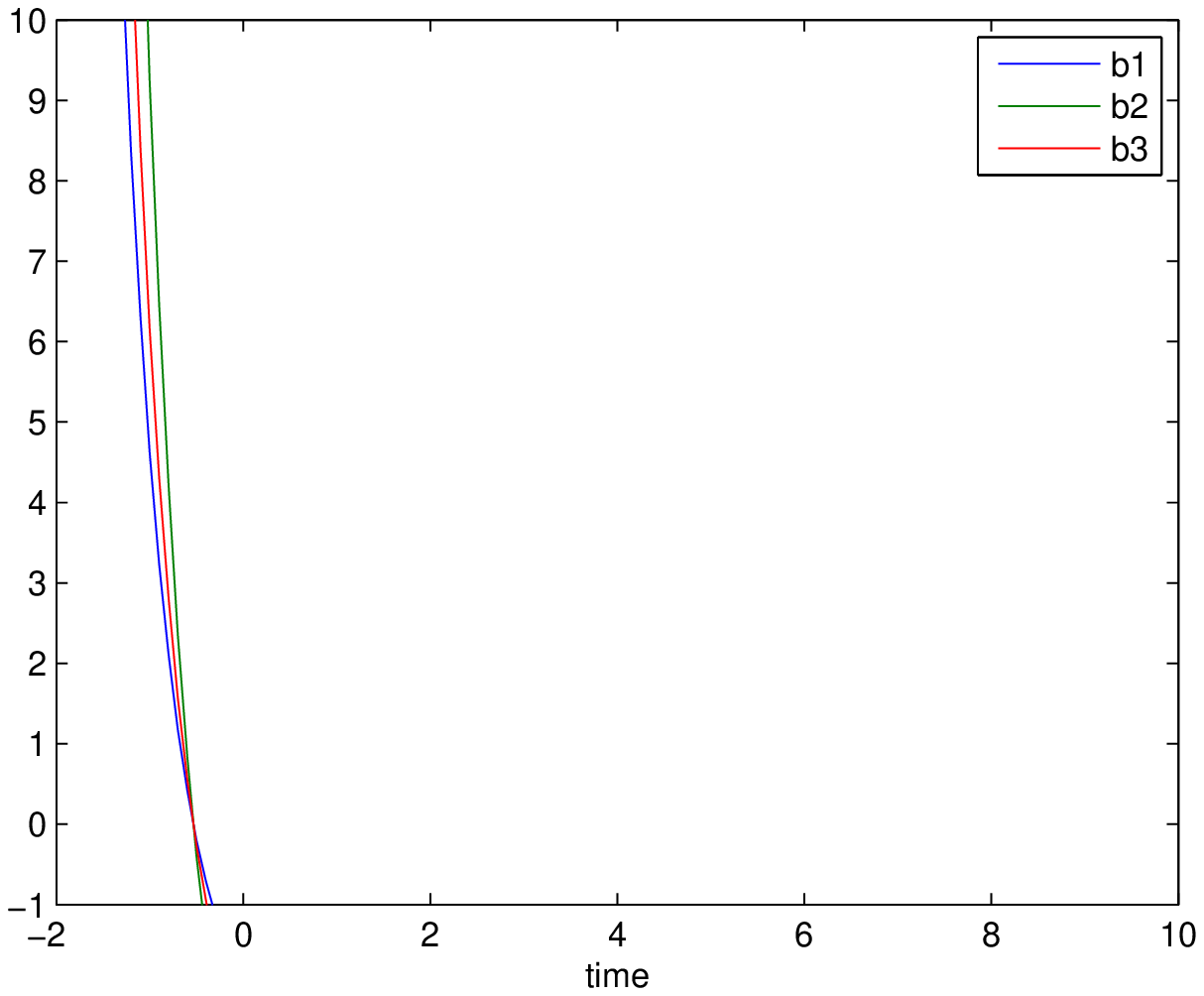}{0.45}{Variation of creation pressure $Pc$ with time
for the subcase Uniform Particle Number Density in Model 2. $b1$,
$b2$ and $b3$ represent the variation for $\gamma=0$, $\gamma=1$ and
$\gamma=1/3$ respectively.} {0.45}{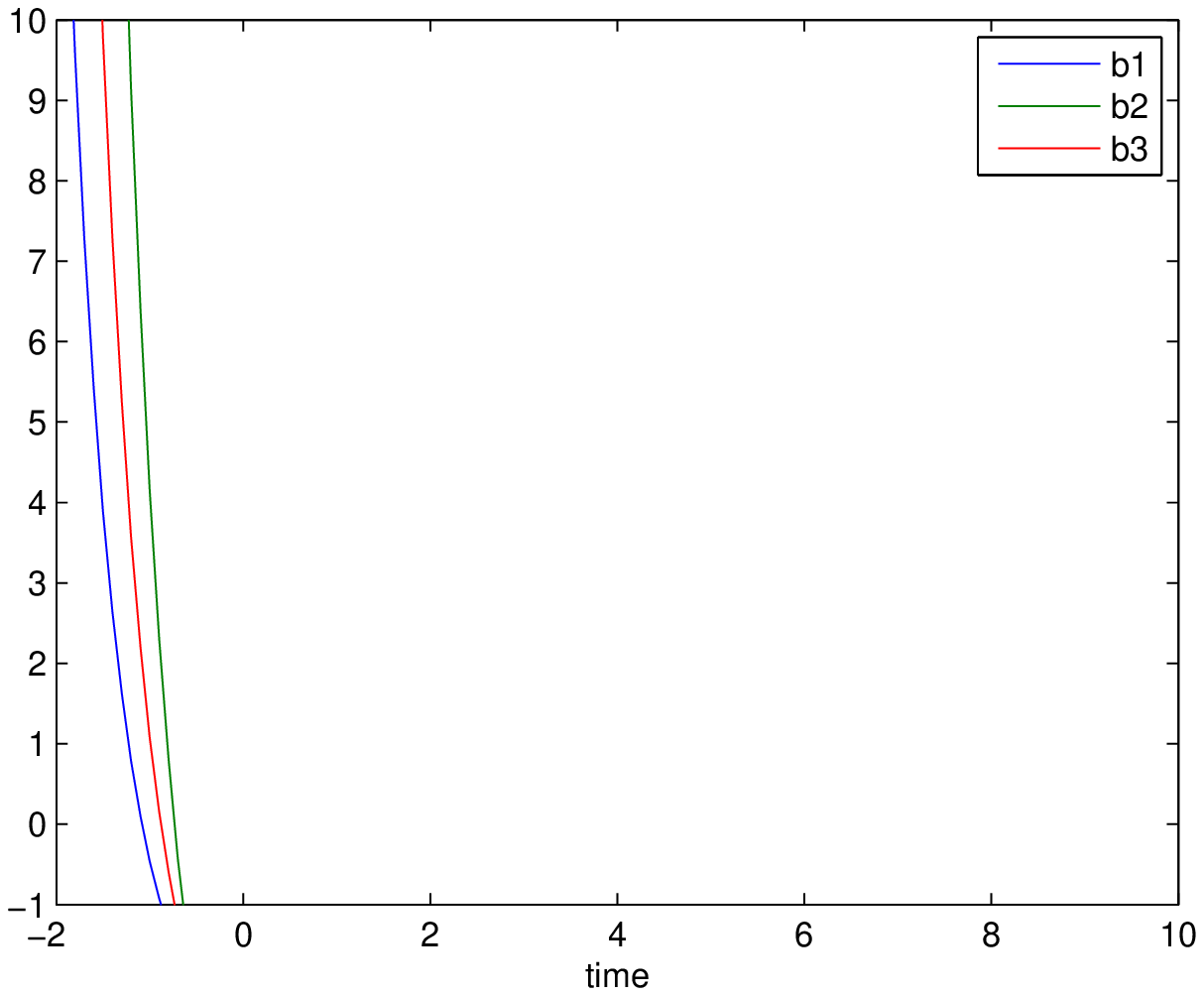}{0.45}{Variation of $\Pi$
with time for the subcase Ideal Gas in Model 2. $b1$, $b2$ and $b3$
represent variation for $\gamma=0$, $\gamma=1$ and $\gamma=1/3$
respectively.}{0.45}

\noindent The coefficient of Bulk viscosity in all three different theories can be calculated as\\

\noindent Truncated theory:\\
\begin{equation}
\zeta= \frac{\displaystyle
\begin{array}{c} [\Pi_1-\Pi_2\exp{(-2h_0t)}][\rho_3-\rho_6 \exp{(-2h_0t)}]
\end{array}}{\displaystyle
\begin{array}{c} \zeta_{71}-\zeta_{72}\exp{(-2h_0t)}
\end{array}}
\end{equation}
where $\zeta_{71}=3 h_0\rho_3$, $\zeta_{72}=(3\rho_6+2\Pi_2)h_0$ .\\

\noindent Full Causal theory:\\
\begin{equation}
\zeta= \frac{\displaystyle
\begin{array}{c} [\Pi_1-\Pi_2\exp{(-2h_0t)}][\rho_3-\rho_6 \exp{(-2h_0t)}]^{2}
\end{array}}{\displaystyle
\begin{array}{c} \zeta_{73}+\zeta_{74}\exp{(-2h_0t)}-\zeta_{75}\exp{(-4h_0t)}
\end{array}}
\end{equation}
where $\zeta_{73}=\rho_3\left(3h_0\rho_3-\frac{3}{2} h_0\Pi_1\right)$,\\
 $\zeta_{74}=\left[\rho_3\left(\frac{3}{2} h_0\Pi_2-2h_0\Pi_2-3h_0\rho_6\right)-\rho_6\left(3h_0\rho_3
 -\frac{3}{2} h_0\Pi_1\right)+\frac{(1+2\gamma)}{(1+\gamma)}h_0 \rho_6\Pi_1\right]$ \\
 and $\zeta_{75}=\left[\rho_6\left(\frac{3}{2} h_0\Pi_2-2h_0\Pi_2-3h_0\rho_6\right)
 +\frac{(1+2\gamma)}{(1+\gamma)}h_0 \rho_6\Pi_2\right]$.\\

\noindent Eckart theory:\\
\begin{equation}
\zeta=\frac{[\Pi_1-\Pi_2\exp{(-2h_0t)}]}{3h_0}.
\end{equation}
\subsubsection{Creation with Second Order Correction in $H$}
Also in this case, the creation pressure and bulk viscous stress can be given as
\begin{equation}
p_c=-\frac{(1+\gamma)d h_0}{3}[\rho_3-\rho_6\exp{(-2h_0t)}]
\end{equation}
\begin{equation}
\Pi=\Pi_3-\Pi_4\exp{(-2h_0t)}
\end{equation}
where $\Pi_3=\frac{d(1+\gamma)h_0\rho_3}{3}-3(1+\gamma)\rho_3 h_0$,
$\Pi_4=\frac{d(1+\gamma)h_0\rho_6}{3}-3(1+\gamma)\rho_6 h_0$. The nature of the graphs in this case for $\Pi$
and $Pc$ can be observed in the figures 31 and 32. \\

\myfigures{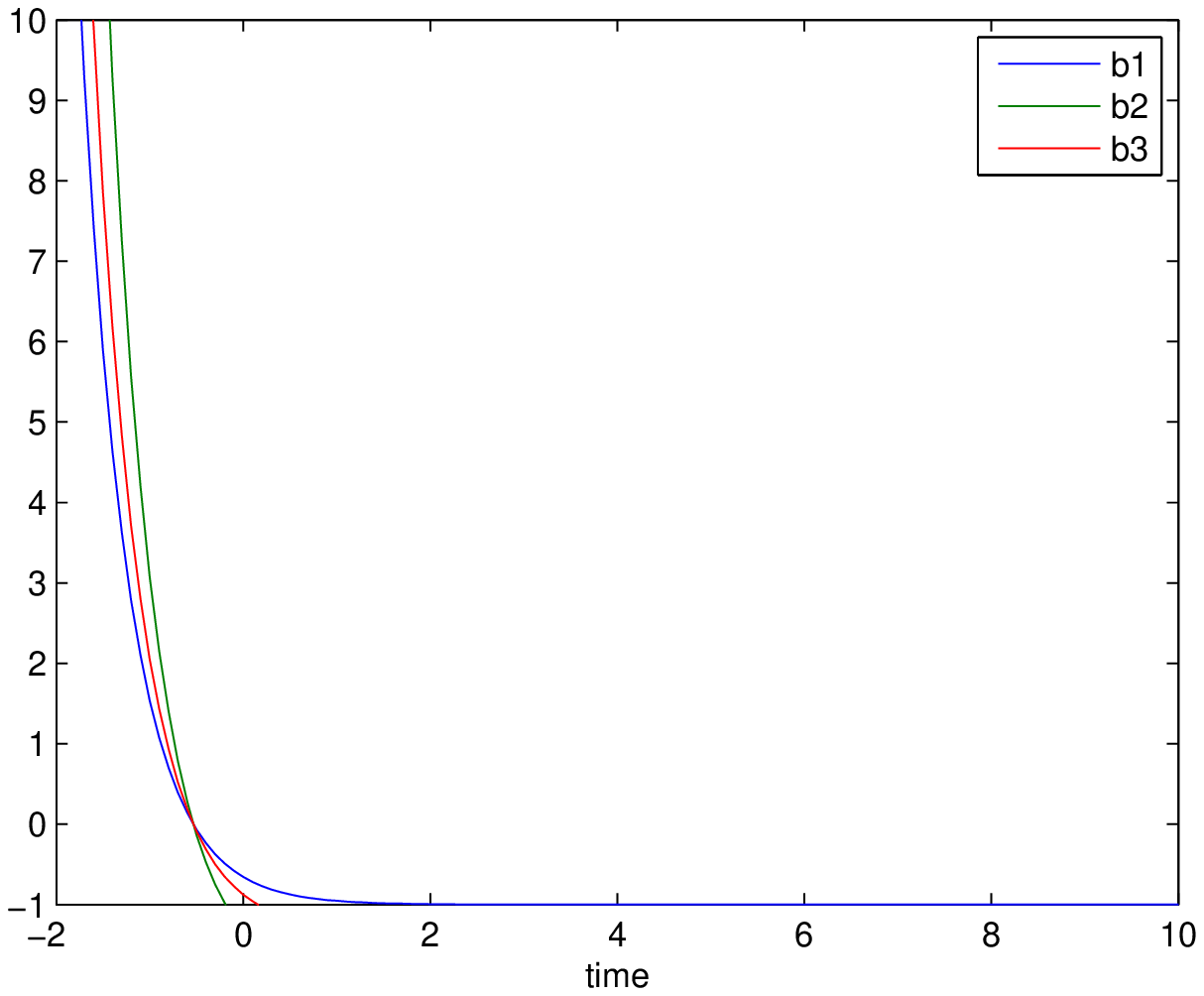}{0.45}{Variation of creation pressure $Pc$ with time
for the subcase Creation with Second Order Correction in $H$ in
Model 2. $b1$, $b2$ and $b3$ represent the variation for $\gamma=0$,
$\gamma=1$ and $\gamma=1/3$ respectively.}
{0.45}{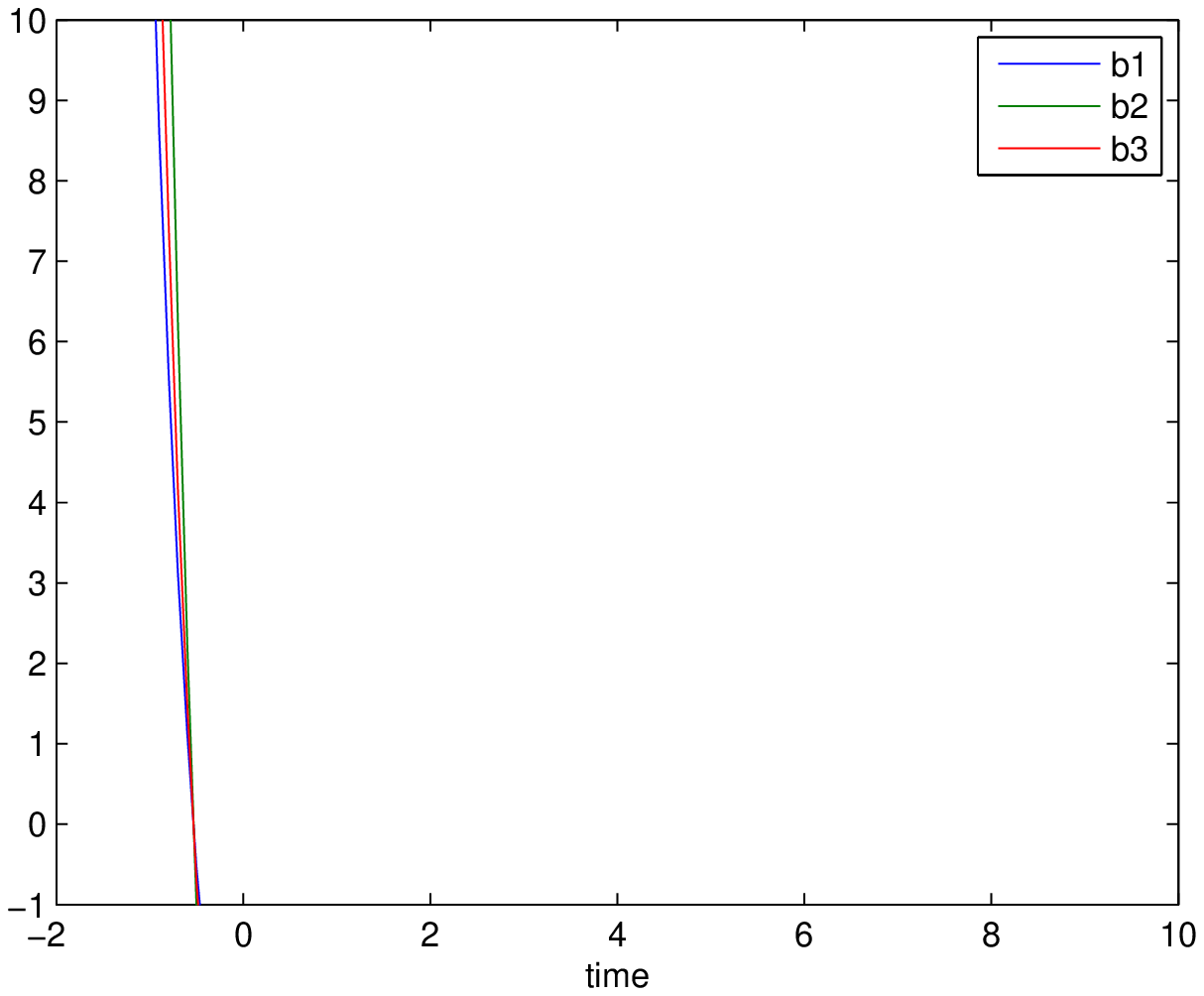}{0.45}{Variation of $\Pi$ with time for the subcase
Creation with Second Order Correction in $H$ in Model 2. $b1$, $b2$
and $b3$ represent the variation for $\gamma=0$, $\gamma=1$ and
$\gamma=1/3$ respectively.}{0.45}

\noindent The bulk viscosity coefficient in all three cases can be obtained \\

\noindent Truncated theory:\\
\begin{equation}
\zeta= \frac{\displaystyle
\begin{array}{c} -[\Pi_3-\Pi_4\exp{(-2h_0t)}][\rho_3-\rho_6\exp{(-2h_0t)}]
\end{array}}{\displaystyle
\begin{array}{c} \zeta_{76}\exp{(-2h_0t)}+3h_0[\rho_3-\rho_6\exp{(-2h_0t)}]
\end{array}}
\end{equation}
where $\zeta_{76}=2h_0 \rho_6\left[\frac{d(1+\gamma)h_0}{3}-3(1+\gamma)h_0-\frac{2}{3}\right]$.\\

\noindent Full Causal theory:\\
\begin{equation}
\zeta= \frac{\displaystyle
\begin{array}{c} -[\Pi_3-\Pi_4\exp{(-2h_0t)}][\rho_3-\rho_6\exp{(-2h_0t)}]^{2}
\end{array}}{\displaystyle
\begin{array}{c} [\rho_3-\rho_6\exp{(-2h_0t)}][\zeta_{77}+\zeta_{78}\exp{(-2h_0t)}-\zeta_{79}\exp{(-4h_0t)}]\\
\end{array}}
\end{equation}
where $\zeta_{77}=\left(3h_0\rho_3+\frac{3}{2}h_0\Pi_3\right)$, $\zeta_{78}=\left[\zeta_{76}-3h_0\rho_6
-\frac{3}{2}h_0\Pi_4+\frac{(1+2\gamma)\rho_6\Pi_3}{2(1+\gamma)}\right]$ and
$\zeta_{79}=\frac{(1+2\gamma)\rho_6\Pi_4}{2(1+\gamma)}$.\\

\noindent Eckart theory:\\
\begin{equation}
\zeta= \frac{\displaystyle
\begin{array}{c} -[\Pi_3-\Pi_4\exp{(-2h_0t)}]
\end{array}}{\displaystyle
\begin{array}{c} 3h_0\\
\end{array}}.
\end{equation}

\section{Conclusion}
In this paper, we have obtained exact solutions for the classical
Einstein's general relativity equation in Bianchi type V space-time
with bulk viscosity in the presence of particle creation. Here, we
have applied the variation law of Hubble's parameter that yields a
constant value of deceleration parameter to find out the solution.
Following this, we have found two types of cosmological models for
two different values of average scale factor obtained from the
assumption. We have found singularity in the first model whereas the
second model is free from it. We have then studied the bulk
viscosity and particle creation in each model for four different
cases. Also we have obtained the bulk viscosity coefficient for
Truncated, Full Causal and Eckart theories in all four cases.
Different physical and kinematical parameters have been obtained and
studied graphically in both the models.

\end{document}